\documentclass[conference]{IEEEtran}
\usepackage{bm}
\usepackage{url}
\usepackage{comment}
\usepackage{graphicx}
\usepackage{amsfonts}
\usepackage{multirow}
\usepackage{tabularx}
\usepackage{booktabs}
\usepackage{xcolor}
\usepackage{color}
\usepackage{algorithm}
\usepackage[noend]{algpseudocode}
\usepackage{adjustbox}
\usepackage{pifont}
\newcommand{\cmark}{\ding{51}}
\newcommand{\xmark}{\ding{55}}
\usepackage{multirow}

\usepackage{etoolbox}
\makeatletter
\patchcmd{\@makecaption}
  {\scshape}
  {}
  {}
  {}
\makeatother

\usepackage{mathtools}
\DeclarePairedDelimiter{\ceil}{\lceil}{\rceil}

\ifCLASSINFOpdf
\else
\fi
\begin{document}
\title{Accelerating ViT Inference on FPGA through Static and Dynamic Pruning}


% \author{\IEEEauthorblockN{}
% \IEEEauthorblockA{School of Electrical and\\Computer Engineering\\
% University of Southern California\\
% Atlanta, Georgia 30332--0250\\
% Email: http://www.michaelshell.org/contact.html}
% \and
% \IEEEauthorblockN{Homer Simpson}
% \IEEEauthorblockA{Twentieth Century Fox\\
% Email: homer@thesimpsons.com}
% \and
% \IEEEauthorblockN{James Kirk\\ and Montgomery Scott}
% \IEEEauthorblockA{Starfleet Academy\\
% San Francisco, California 96678--2391\\
% Telephone: (800) 555--1212\\

\author{
\IEEEauthorblockN{Dhruv Parikh\IEEEauthorrefmark{1}, Shouyi Li\IEEEauthorrefmark{1}, Bingyi Zhang\IEEEauthorrefmark{1}, Rajgopal Kannan\IEEEauthorrefmark{2},  Carl Busart\IEEEauthorrefmark{2}, Viktor Prasanna\IEEEauthorrefmark{1}}
\IEEEauthorblockA{
    \IEEEauthorrefmark{1}University of Southern California \IEEEauthorrefmark{2}DEVCOM Army Research Office \\
    \IEEEauthorrefmark{1}\{dhruvash, liderric, bingyizh, prasanna\}@usc.edu \IEEEauthorrefmark{2}\{rajgopal.kannan.civ, carl.e.busart.civ\}@army.mil}
}


%Telephone: (800) 555--1212, Fax: (888) 555--1212}


\maketitle


% To do: make concise
\begin{abstract}

Vision Transformers (ViTs) have achieved state-of-the-art accuracy on various computer vision tasks. However, their high computational complexity prevents them from being applied to many real-world applications. Weight and token pruning are two well-known methods for reducing computational complexity. Weight pruning reduces the model size and associated computational demands, while token pruning further reduces the computation based on the input.
Combining these two techniques should significantly reduce computation complexity and model size; however, naively integrating them results in irregular computation patterns, leading to significant accuracy drops and difficulties in hardware acceleration.

To address the above challenges, we propose a comprehensive algorithm-hardware codesign for accelerating ViT on FPGA through {\it simultaneous pruning} -- combining \emph{static} weight pruning and \emph{dynamic} token pruning. For \emph{algorithm} design, we systematically combine a hardware-aware structured block-pruning method for pruning model parameters and a dynamic token pruning method for removing unimportant token vectors. Moreover, we design a novel training algorithm to recover the model's accuracy. For \emph{hardware} design, we develop a novel hardware accelerator for executing the pruned model. The proposed hardware design employs multi-level parallelism with a load-balancing strategy to efficiently deal with the irregular computation pattern led by the two pruning approaches. Moreover, we develop an efficient hardware mechanism for executing the on-the-fly token pruning. We apply our codesign approach to the widely used DeiT-Small model. We implement the proposed accelerator on a state-of-the-art FPGA board. The evaluation results show that the proposed algorithm can reduce computation complexity by up to $3.4\times$ with $\approx 3\%$ accuracy drop and a model compression ratio of up to $1.6\times$. Compared with state-of-the-art implementation on CPU, GPU, and FPGA, our codesign on FPGA achieves an average latency reduction of $12.8\times$, $3.2\times$, and $0.7-2.1\times$, respectively.

\end{abstract}
% make more precise


% \fi
\IEEEpeerreviewmaketitle

\section{Introduction}
\label{sec: 1_intro}

Vision Transformers (ViTs) \cite{dosovitskiy2021image}  have demonstrated superior performance in comparison to Convolutional Neural Networks (CNNs) in various vision tasks \cite{NEURIPS2021_854d9fca, pmlr-v119-chen20s, chen2021empirical, carion2020endtoend, DBLP:journals/corr/abs-2012-00759/maxdeep, jiang2021transgan, ramesh2021zeroshot}. The global self-attention in ViTs leads to a reduced local and image-specific inductive bias \cite{dosovitskiy2021image}; this results in ViTs requiring larger datasets and larger model sizes \cite{Khan_2022} to perform better than CNN. The Multi-head Self-Attention (MSA) of ViTs allows them to generalize better than CNNs on larger datasets \cite{park2022vision}. However, their computational cost is usually significantly higher than CNNs due to the MSA mechanism, which scales quadratically with the number of input tokens \cite{zhu2024token, vaswani2023attention}. Their intensive computational requirements emphasize the need for efficient hardware acceleration. 

In addressing the computational challenge, pruning has been proven to be effective in reducing the computational cost of CNNs \cite{ma2021sanity, frankle2019lottery, liu2021lottery, zhang2020unified}. However, explorations in self-attention-based pruning methods still need to be discovered \cite{sanh2020movement, li2020efficient, wang2021spatten}. Many existing works on efficient ViTs explored block weight pruning and token pruning as two distinct strategies. Weight pruning, introduced in \cite{anwar2015structured, molchanov2017pruning, He_2023_cnnprune_survey, DBLP:journals/corr/abs-2109-04838_blk_prn_fstr_xmer, sanh2020movement, han2015learning_pruning, li2020efficient, Yu_Huang_Wang_Cheng_Chu_Cui_2022, Yu_Huang_Wang_Cheng_Chu_Cui_2022, NEURIPS2022_3b11c5cc_savit}, reduces the model size by pruning input parameters statically and selectively, thus feeding the neural network with sparse inputs to reduce computation. Token pruning removes tokens to reduce the computational complexity. The static approaches in \cite{chen2021chasing, liang2022patches, fayyaz2022adaptive, tang2022patch} drop tokens with a fixed ratio, often ignoring the redundancies between tokens; dynamic token pruning studies in \cite{pan2021iared2, xu2021evovit, yu2022unified} do not fully explore the token redundancies from different attention heads and simply discard non-informative tokens. \cite{kim2022learned, kong2022spvit, rao2021dynamicvit, dong2023heatvit} dynamically reduce the number of tokens in ViT encoders during inference based on the inherent image characteristics. Moreover, only a few of these studies support efficient hardware implementations by the respective pruning algorithm. Both weight pruning and token pruning methods reduce the computational complexity independently, but the interaction between the two remains unexplored. A combined approach could bring further computational benefits. However, such integration poses two main challenges: (1) accuracy drop (algorithm level) and (2) increased computational pattern irregularities (hardware level).

% \textcolor{red}{[notes]: Fixing the intro: 
% * Be more specific and expanded on FPGA part (e.g. Latest FPGA platform/resources - better at handle irregular patterns)
% * CPU and GPU are not efficient -> FPGA is the way to go
% * Algorithm advantages -> but need to implement it on FPGA to achieve high efficiency and explain WHY: two pruning techniques (1) weight pruning produces sparse matrix inputs (2) token pruning requires token shuffling, CPU and GPU does not have good capabilities to handle this (3) work-load imbalance: CPU and GPU needs complicated process, FPGA is better to balance the work-load)
% }

% \textcolor{red}{[comments]:  Write a paragraph to describe/explain why existing FPGA accelerator cannot efficiently execute the two pruning approaches. For example, we can say some FPGA accelerator can only support weight pruning, some FPGA accelerator only support token pruning. Why CPU and GPU not efficient}

Many ViT acceleration works primarily focus on the CPU and GPU platforms \cite{liang2022patches, rao2021dynamicvit, xu2021evovit, fayyaz2022adaptive, tang2022patch}. However, the integration of block weight pruning and token pruning in ViTs effectively reduces the model size, thus making it possible to accommodate the compressed model onto FPGA. 
% As an alternative to GPUs, FPGAs, with their low latency, energy efficiency, and reconfigurability, become increasingly popular and are widely used for machine learning models and real-time processing \cite{KuonRose2007, guo2018survey, guan2017fpga}. 
%While existing ViT  accelerators on FPGA \cite{dong2023heatvit, kong2022spvit} can handle the irregular patterns after pruning, they target either weight pruning or token pruning, but not both. Therefore, none of the existing FPGA ViT accelerators can support our integrated approach (simultaneous pruning). 
Comprehensively, we use FPGA to accelerate our pruned ViT models for these reasons: (1) %Although CPUs and GPUs are capable of processing the sparse matrix inputs produced by block weight pruning, FPGAs, with customized data path and on-chip memory organization, stand out as better choices to maximize the computation efficiency. 
FPGAs, with customized data path and on-chip memory organization, stand out as better choices than CPU/GPU to maximize the computation efficiency. 
(2) CPUs and GPUs cannot effectively handle the token shuffling process of our dynamic token pruning. We design a specific FPGA kernel to handle the token shuffling in the middle of model inferences. (3) CPUs and GPUs need complicated processes to address work-load imbalance, whereas on FPGA, we can design customized hardware modules for balancing work-load.

In this paper, we propose an algorithm-hardware codesign for accelerating ViT inference. Different from existing ViT acceleration works, we utilize the combined power of \emph{static} weight pruning and \emph{dynamic} token pruning. We propose a simultaneous pruning algorithm to recover the model accuracy caused by two pruning approaches. Combining the two pruning approaches leads to more severe computational irregularity. Therefore, we develop a customized data path and memory organization on FPGA to execute the pruned model efficiently. 
While existing ViT accelerators on FPGA \cite{kong2022spvit, dong2023heatvit} can handle the irregular patterns after pruning, they target either weight pruning or token pruning, but not both. Therefore, none of the existing FPGA ViT accelerators can support our integrated simultaneous pruning approach.
We summarize our main contributions below:

% \textcolor{red}{[comments]: This paragraph should briefly talk about how our work address the two challenges of combing two pruning algorith: (1) accuray drop, (2) irregular computation pattern, through algorithm-hardware codesign.}

\begin{itemize}

    \item We propose an algorithm-hardware codesign for efficient ViT inference based on FPGA. The design combines parameter (static) and token (dynamic) pruning to reduce both the ViT model size and computational complexity.

    \item For the algorithm design, we systematically combine \emph{static} block weight pruning and \emph{dynamic} token pruning to reduce the computation complexity of ViTs. We propose a novel training algorithm to recover the accuracy drop led by the two pruning algorithms.
    
    % \item The proposed framework reduces the model size by performing block-wise pruning of ViT model parameters during the fine-tuning phase; this block-wise pruning is done offline, statically to make the encoder parameters sparse. The pruning is hardware-friendly; during this phase, we remove redundant attention heads from MSA and block-wise prune the remaining heads, additionally, we remove redundant neurons from each layer in the encoder multi-layer perceptron (MLP).

    % \item The proposed accelerator, further reduces the computational complexity for model inference by dynamically pruning the inattentive input tokens during inference; the token keep-ratio decides the fraction of tokens that will be retained at a specific stage. Our dynamic token pruning removes tokens within an encoder, prior to the MLP segment, leading to an intra and inter encoder reduction in computations.
    \item For the hardware design, we develop a novel hardware accelerator with multi-level parallelism and a load balancing strategy. This can efficiently deal with (1) load imbalance caused by the block pruning and (2) a changing number of tokens caused by token pruning. We also develop an efficient hardware mechanism for executing the on-the-fly token pruning algorithm.

    \item We evaluate our codesign on DeiT models and deploy the proposed accelerator on a state-of-the-art FPGA board - Xilinx Alveo U250. The evaluation results show that the proposed algorithm can reduce computation complexity by up to $3.4\times$ with $\approx 3\%$ accuracy drop and a model compression ratio of up to $1.6\times$. Compared with state-of-the-art implementation on CPU, GPU, and FPGA, our codesign on FPGA achieves average latency reduction of $12.8\times$, $3.2\times$, $0.7-2.1\times$ respectively.
    
\end{itemize}

% To the best of our knowledge, our end-to-end framework is the first of its kind that combines static (offline) and dynamic (online) pruning to optimize model inference for ViTs via a customized accelerator, generated end-to-end, that exploits sparsity along both the model size and input tokens. 

% To the best of our knowledge, the proposed work is the first algorithm-hardware codesign on FPGA that combining the static weight pruning and dynamic token pruning for accelerating ViTs. This paper is organized as follows, Section \ref{sec:2_background_related_work} introduces the background and related works; Section \ref{sec: proposed_method} introduces the overview. Section \ref{sec: exp_res} covers the performed experiments with our framework and the associated results and comparisons. Finally, section \ref{sec: concl}, \ref{sec: disc}, \ref{sec: future} conclude the paper with concluding discussions and potential for future work.

%%%%% [?; required addition] [Details] FPGA: context, FPGA: our novelty %%%%%%

\section{Background and Related Work}
\label{sec:2_background_related_work}

%%%% COMMENTS, DP %%%%%
%%%% Go through revised sections and correct some stuff %%%%
%%%%%%%%%%%%%%%%%%%%%%%

\subsection{Vision Transformer}

ViT \cite{dosovitskiy2021image} has a stack of transformer encoders. Each encoder has a multi-head self-attention (MSA) and a multi-layer perceptron (MLP). The input image $\mathbf{x} \in \mathbb{R}^{H \times W \times C}$ is first partitioned into $N$ patches $\mathbf{x}_p \in \mathbb{R}^{N \times P^{2} C}$. 
% Each patch has a resolution of $(P, P)$ within the original image of resolution $(H, W)$ and the number of patches, $N$, is $HW/P^2$, with each patch flattened into a vector.
Each patch is flattened into a vector of length $P^{2}C$.
Next, a learnable linear mapping  method 
%$\mathbf{E} \in \mathbb{R}^{P^2 C \times D}$ 
maps each patch to a token vector of length $D$. A special parameterized token $\mathbf{x}_{\text{CLS}}$ is appended as a token vector. Then, a positional embedding $\mathbf{E}_{\text{POS}} \in \mathbb{R}^{(N+1) \times D}$ is added to input token matrix to produce $\mathbf{Z_0} \in \mathbb{R}^{(N+1) \times D}$ which is the input to the transformer encoder stack. 
% The linear mapping $\mathbf{E}$ is generally realized via a 2D convolution kernel of size $(P, P)$, input channels $C$, output channels $D$ and stride $(P, P)$. 
For simplicity, we denote the number of input tokens to the encoder stack as $N$ instead of $N+1$ for the rest of the paper.

% The input $\mathbf{z_0}$ passes through each encoder in the encoder stack sequentially; the operations within an encoder $l$, can be described in two stages, as below:

% (to save space, this figure can be removed since we have all the equation)
% \begin{figure}[h]
% \includegraphics[width=0.45\textwidth]{figures/vitFig.pdf}
% \centering
% \caption{Diagram of a ViT encoder}
% \label{fig: encoder_basic}
% \end{figure}

% \begin{figure}[h]
% \includegraphics[width=0.45\textwidth]{figures/encoder_basic.jpg}
% \centering
% \caption{Diagram of ViT encoder stack; the encoder at depth $l$ and the associated MSA and MLP segments have been highlighted}
% \label{fig: encoder_basic}
% \end{figure}

% MSA stage

\textbf{MSA.} The input to encoder $\mathbf{Z}_{l-1}$, is layer normalized (LN) \cite{ba2016layer} and passed through a multi-headed self-attention (MSA) layer \cite{vaswani2023attention}:
\begin{equation}
    \mathbf{Z}_{l}{'} = \mbox{MSA}(\mbox{LN}(\mathbf{Z}_{l-1})) + \mathbf{Z}_{l-1}
\label{eq: MSA stage}
\end{equation}
$\mbox{MSA}(\cdot)$ is expressed as: 
\begin{equation}
\begin{split}
    [\mathbf{Q}, \mathbf{K}, \mathbf{V}] = \mathbf{Z}\mathbf{U}_{qkv}
\end{split}
\label{eq: q,k,v gen}
\end{equation}
where $\mathbf{U}_{qkv}=[\mathbf{W}_{q}, \mathbf{W}_{k}, \mathbf{W}_{v}] \in \mathbb{R}^{D \times 3D'} $, $\mathbf{Z} \in \mathbb{R}^{N \times D}$, $\mathbf{Q}, \mathbf{K}, \mathbf{V} \in \mathbb{R}^{N \times D'}$, corresponding to \emph{query}, \emph{key} and \emph{value} matrices, respectively. $D$ is the length of input token and $D'$ is the hidden dimension. Then, the attention score matrix $\mathbf{A}$ is calculated through:
% he query of each token computes a dot-product attention score with respect to keys of all the tokens; the scores are then used to compute a weighted aggregate of values to obtain the self-attention output for a particular token (across all tokens).  
\begin{equation}
    \mathbf{A} = \mbox{softmax}(\frac{\mathbf{Q} \mathbf{K}^{T}}{\sqrt{D'}}) \quad \mbox{where} \quad \mathbf{A} \in \mathbb{R}^{N \times N}
\label{eq: attention}
\end{equation}
\begin{equation}
    \mbox{SA}(\mathbf{Z}) = \mathbf{A} \mathbf{V} \quad \mbox{where} \quad \mbox{SA}(\mathbf{Z}) \in \mathbb{R}^{N \times D'}
\label{eq: self atten}
\end{equation}
$\mbox{SA}(\cdot)$ refers to self-attention with a single head. $\mbox{MSA}(.)$ extends this notion of self-attention to several parallel heads, each with its own parameters:
% computing on the same input $\mathbf{z}$ (in parallel).
\begin{equation}
    \mbox{MSA}(\mathbf{Z}) = [\mbox{SA}_{1}(\mathbf{Z}) \quad \mbox{SA}_{2}(\mathbf{Z}) \quad \cdots \quad \mbox{SA}_{H}(\mathbf{Z})] \mathbf{W}_{\text{proj}} 
\label{eq: msa}
\end{equation}
where $H$ denotes the total number of heads. $\mathbf{W}_{\text{proj}}  \in \mathbb{R}^{HD' \times D}$ projects the concatenated self-attention outputs of the individual heads back to the embedding dimension $D$.

% MLP stage

\textbf{MLP.} The output of MSA, $\mathbf{Z}_{l}{'}$ is layer normalized and passed through a multi-layer perceptron (MLP):
\begin{equation}
    \mathbf{Z}_{l} = \mbox{MLP}(\mbox{LN}(\mathbf{Z}_{l}{'})) + \mathbf{Z}_{l}{'}
\label{eq: mlp}
\end{equation}

% The $\mbox{MLP}(.)$ in eq. \ref{eq: mlp} is a standard feed-forward neural network (FFN), typically with two layers and a GELU activation \cite{hendrycks2023gaussian}. The first (intermediate) layer expands the embedding dimension from $D$ to $D_{mlp}$ and the second (output) layer contracts it back to $D$. Thus, $\mathbf{z}_{l} \in \mathbb{R}^{N \times D}$.

% For image classification, the state of the class token at the final encoder ($\mathbf{z}_{L}^{0}$) is utilized as the learned feature representation of the input image. An MLP head attached to this token yields the final class logits.

% \begin{equation}
%     \mbox{logits}_{\mathbf{x}} = \mbox{MLP}_{cls}(\mbox{LN}(\mathbf{z}_{L}^{0}))
% \label{eq: cls mlp head}
% \end{equation}

% In eq. \ref{eq: cls mlp head}, $\mbox{logits}_{\mathbf{x}} \in \mathbb{R}^{c}$ represent the logits of input image $\mathbf{x}$ with total classes $c$ in the dataset. $\mbox{MLP}_{cls}$ is typically just a single layer FFN.  

\subsection{Related Work}
\label{subsec: rel_work}

\noindent \textbf{Weight pruning}: Structured and hardware-friendly model parameter pruning, used in traditional CNNs \cite{anwar2015structured, molchanov2017pruning, He_2023_cnnprune_survey}, becomes popular for ViT. \cite{sanh2020movement} introduces the notion of movement pruning, which prunes parameters by generating a pruning mask based on the learned scores.
\cite{DBLP:journals/corr/abs-2109-04838_blk_prn_fstr_xmer} proposes to prune parameters across all the encoders. Magnitude-based approaches, on the other hand, discard parameters with large magnitudes \cite{han2015learning_pruning}. \cite{li2020efficient} partitions a parameter matrix into blocks and prunes the rows and columns in each block by using the $l_2$ norms. \cite{Yu_Huang_Wang_Cheng_Chu_Cui_2022} prunes the entire attention heads within the MSA and neurons in each feed-forward linear layer (width pruning). \cite{Yu_Huang_Wang_Cheng_Chu_Cui_2022} also removes entire encoders after a certain depth (depth pruning). \cite{NEURIPS2022_3b11c5cc_savit} proposes a collaborative approach to optimizing ViT pruning that prunes heads and neurons; the latter neurons are pruned such that they reduce the length of each token. This method considers the collective pruning impact through an expensive approximation of the Hessian of the loss.

\noindent \textbf{Token pruning}: 
% Besides structured model parameter pruning, token pruning approaches have recently been a research focus. The computation complexity of each compute operation within an encoder has at least a linear dependency on the number of tokens, with certain operations having a squared dependency (see section \ref{sec: Overview});
Token pruning approaches attempt to identify redundant tokens and drop them to reduce the computational footprint associated with the number of tokens. Both \cite{wang2021spatten} and \cite{lu2021sanger} have been notable for accelerating transformer models by leveraging the inherent sparsity in attention mechanisms. However, they do not use weight and token pruning simultaneously. \cite{liang2022patches} proposes a static approach to token dropping that ranks the importance of tokens by the attention score of the class token with respect to each token aggregated across heads. In theory, such static approaches do not need additional training since the token-dropping module is not parameterized. In contrast to static token pruning, dynamic token pruning as employed in \cite{rao2021dynamicvit, kong2022spvit, kim2022learned} adds additional model parameters that facilitate the selection of relevant/attentive tokens. \cite{rao2021dynamicvit} and \cite{kong2022spvit} utilize a token selector network inserted at various depths along the original transformer network; such token selector networks are neural networks that output a decision (binary) mask to inform the retention or removal of a token. \cite{kim2022learned}, on the other hand, associate a learnable score to each token and prune tokens with scores lower than a threshold.

% \vspace{0.1cm}
% \noindent \textbf{Hardware acceleration of ViTs}: Transformer acceleration paradigms driven by model compression and/or computational complexity reduction (optimization) have drawn significant attention from research community of late \cite{huang2022hardware} due to the massive potential displayed by transformers in applications encompassing a wide variety of domains \cite{islam2023comprehensive}. \cite{9424344} performs structured block pruning on transformer network parameters and enforces balance in each column-block across the column-blocks of a particular parameter matrix. A specialized format is proposed for the storage of the resultant column-balanced block-pruned sparse matrices and accelerated via a dedicated processing element (PE); the tasks within an encoder/decoder are each assigned to a PE with the FPGA resource allocation constrained via a resource scheduler. 

\section{Overview}
\label{sec: Overview}

%\textbf{1 page for overview}

% Should include computational complexity analysis (theoretical) (baseline)
% include naive computational complexity details

% Add internal compute diagram for MSA and MLP

\subsection{Problem Definition}
\label{sec: 0.1_problem_def}

Our objective is to accelerate ViT on FPGA  through an algorithm-hardware codesign that 1) utilizes a novel combination of model weight pruning and token pruning to reduce computation complexity (algorithm design), and 2) an efficient accelerator that explicitly accounts for the distinct and irregular access patterns of the two pruning techniques for executing the combined pruned model (hardware design).
%Our objective is to accelerate ViT on an FPGA platform  through an algorithm-hardware codesign. For algorithm design, we intend to utilize the combination of model weight pruning and token pruning to reduce the computation complexity. For hardware design, we intend to develop an efficient accelerator for executing the pruned model.

For the algorithm design, the \emph{input} is a ViT model denoted as $\mathcal{M}(\cdot, h_{\text{structure}}, \Theta)$ where $h_{\text{structure}}$ are the model hyper-parameters, including the number of encoders, number of heads, dimensions of linear layers, etc. $\Theta$ is the trainable parameters containing the weights and biases of the MSA and MLP. Our algorithm design prunes the input model $\mathcal{M}$ through (1) \emph{offline} weight pruning that reduces the redundant parameters in $\Theta$, and (2) \emph{runtime} token pruning that trims the number of tokens (in $h_{\text{structure}}$) in the intermediate layers according to the importance of the token. After pruning, we obtain the pruned model denoted as $\mathcal{M}^{'}(\cdot, h_{\text{structure}}^{'}, \Theta^{'})$, where $\Theta^{'}$ denotes the model parameters after weight pruning and $h_{\text{structure}}^{'}$ denotes the hyperparameters after token pruning. Our algorithm design aims to reduce the number of parameters, reduce the computation complexity, and maintain accuracy.

For the hardware design, the hardware accelerator executes the pruned model $\mathcal{M}^{'}(\cdot, h_{\text{structure}}^{'}, \Theta^{'})$. For executing the model, the input is an image sample $\bm{x}$, and the accelerator executes $\mathcal{M}^{'}(\bm{x}, h_{\text{structure}}^{'}, \Theta^{'})$ to obtain the result. The \emph{latency} is the duration from the time when the accelerator receives the input $\bm{x}$ to the time when the accelerator obtains the result.

\subsection{Computational Complexity}

% The computation process of MSA and MLP is demonstrated in Figure \ref{fig: encoder_basic}, respectively.  
The computational complexity for each operation within the MSA and MLP without pruning are summarized in table \ref{tab:complexity_generic}. $B$ denotes the batch size. 

% Add comment regarding MSA block vs MSA operation and MLP block vs MLP operation (inside the blocks)

% \begin{figure}[h]
% \includegraphics[width=0.45\textwidth]{figures/msa_unrolled.jpg}
% \centering
% \caption{Diagram of MSA. Note that MSA is used to refer to both the MSA block (left) and the MSA operation (right). In red, we highlight the outputs associated with a single head.}
% \label{fig: msa_unrolled}
% \end{figure}

% \begin{figure}[h]
% \includegraphics[width=0.45\textwidth]{figures/mlp_unrolled.jpg}
% \centering
% \caption{Diagram of MLP. Note that MLP is used to refer to both the MLP block (left) and the MLP operation (right). The intermediate linear layer has GELU activation and the output linear layer has no activation.}
% \label{fig: mlp_unrolled}
% \end{figure}

\begin{table}[h]
    \centering
    \caption{Computational complexity within a single ViT encoder. $()$ indicates the number of instances inside a single encoder.}
    \vspace{-0.3cm}
    \begin{adjustbox}{max width=0.3\textwidth}
    \begin{tabular}{ cc }
         \toprule
         \textbf{Operation} & \textbf{Computational Complexity} \\
         \midrule
         \midrule
         LayerNorm $(\times 2)$ & $BND$ \\
         \midrule
         Residual Add $(\times 2)$ & $BND$ \\
         \midrule
         MSA  $(\times 1)$ & $4BHNDD' + 2BHN^2D'$ \\
         \midrule
         MLP  $(\times 1)$ & $2BNDD_{\text{mlp}}$ \\
         \midrule
         \textbf{Total Complexity} &\begin{tabular}[|c|]{@{}c@{}} $4BND + 4BHNDD' $ \\ $+ 2BHN^2D' + 2BNDD_{\text{mlp}}$ \end{tabular} \\
         \bottomrule
    \end{tabular}
    \end{adjustbox}
    \vspace{-0.4cm}
    \label{tab:complexity_generic}
\end{table}

\subsection{Overview of Algorithm-hardware Codesign}

\begin{figure}[h]
\includegraphics[width=0.42\textwidth]{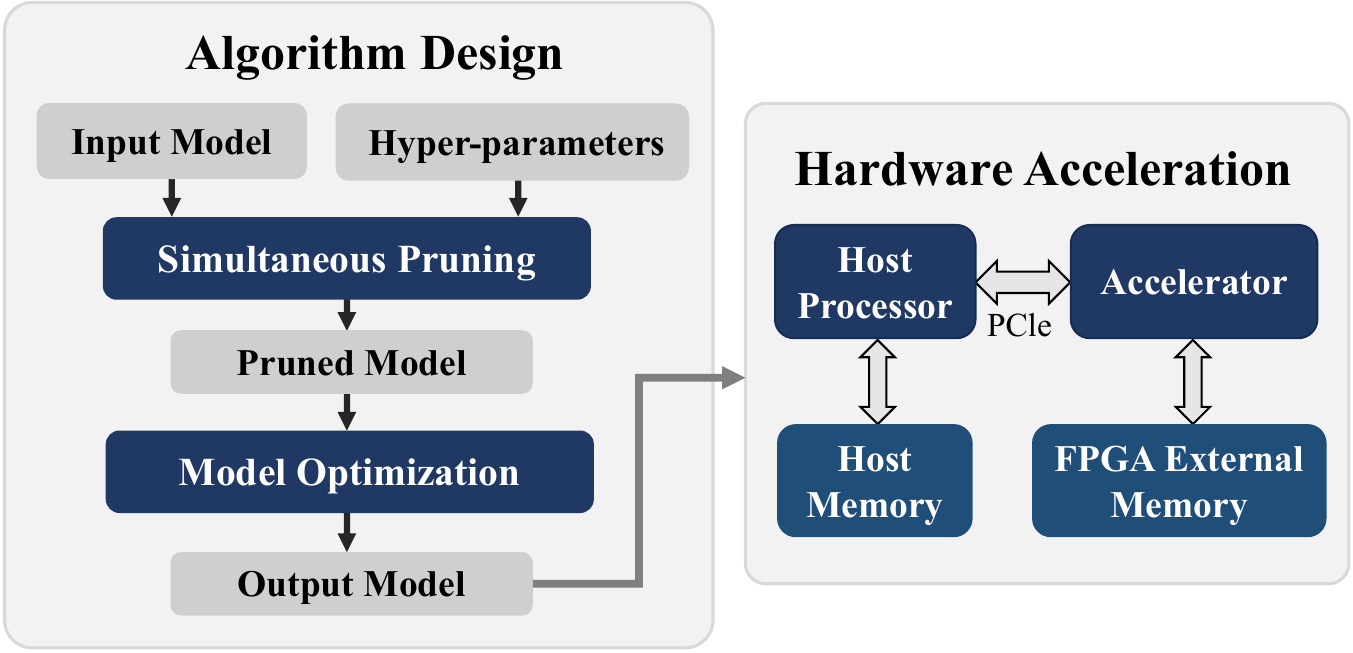}
\centering
\vspace{-0.3cm}
\caption{Overview of the proposed algorithm-hardware codesign}
\label{fig: system_overview}
\vspace{-0.4cm}
\end{figure}

The overview of the proposed codesign is shown in Figure \ref{fig: system_overview}, which consists of algorithm design and hardware design.

\emph{Algorithm design}: For algorithm design, we utilize the combination of block-wise static weight pruning (Section \ref{subsec:model-pruning}) and input token pruning (Section \ref{subsec: dyn_tkn_prn}). The input is the input ViT model and the pruning hyper-parameters (including the weight pruning ratio for each layer and token pruning ratio for each layer). Users manually specify the pruning hyper-parameters. Then, the proposed simultaneous pruning (training) algorithm (Section \ref{subsec:simu-pruning}) prunes the input model according to the user-specified pruning hyper-parameters. Then, the pruned model is generated. The model optimizations organize the data blocks in the weight matrices into the required data layout and format (Section \ref{subsec: hw_data_org}) for efficient hardware execution on the proposed accelerator.

\emph{Hardware accelerator design}: At runtime, when the host process receives an input image, it sends the input image to the accelerator for inference. We employ a multi-level parallelism strategy for the proposed hardware architecture to efficiently handle the irregular computation patterns. We design a Token Dropping Hardware Module for efficient on-the-fly token dropping. See Section (Section \ref{subsec: hw_wrkflw}) for details. 

\vspace{0.1cm}
\noindent \textbf{Discussion on the challenges of hardware acceleration:} The combination of two pruning approaches leads to significant challenges for hardware accelerations: (1) Through weight pruning, the weight matrix of MSA has uneven number of data blocks among different columns and different layers can have different number of heads. Moreover, the token pruning leads to varying number of tokens for different layers. These potentially leads to runtime resource underutilization. We address this challenge through multi-level parallelism (Section \ref{subsec: hw_wrkflw}) with load balance strategy (Section \ref{subsec: opt}). (2) Due to the block-wise weight pruning, both token matrix and weight matrices are partitioned into data blocks. However, the token dropping algorithms drops unimportant tokens in the intermediate layers. Therefore, the token matrix needs to be reordered and reconstructed on the fly based on their importance score. This involves sorting and data shuffling which cannot be efficiently handled by CPU or GPU. We develop efficiently hardware mechanism in Token Dropping Hardware Module to address the above issue (Section \ref{subsubsec:Dynamic-Token-Dropping}).

% Need some hardware details fixed to continue this

\section{Pruning Algorithm}
\label{sec:Pruning}

%%%%%% COMMENTS FOR DP %%%%%%%%%%%
%%%%%% Modifying and make uniform slicing notations %%%%%%%%
%%%%%%%%%%%%%%%%%%%%%%%%%%%%%%%%%%

Existing works only utilize either weight pruning or token pruning. In contrast, we systematically combine the two pruning approaches with a novel training algorithm for recovering accuracy. To this end, we first introduce static weight pruning (Section \ref{subsec:model-pruning}) and dynamic token pruning (Section \ref{subsec: dyn_tkn_prn}) separately, and then introduce our Simultaneous Pruning algorithm to prune the input model.

%\textbf{1-1.5 page for Pruning Algorithm}

% Including structure of pruned layers

% Including baseline computational complexity with pruning

\subsection{Static Weight Pruning}
\label{subsec:model-pruning}

% We utilize a score movement based block pruning approach \cite{9424344, DBLP:journals/corr/abs-2109-04838_blk_prn_fstr_xmer, sanh2020movement}. 
The weights to be pruned are: weight matrices for $\mathbf{W}_{q}$, $\mathbf{W}_{k}$, $\mathbf{W}_{v}$ and $\mathbf{W}_{\text{proj}}$ within MSA and the \emph{intermediate} and \emph{output} linear layers within MLP. Pruning is performed as follows:

\subsubsection{Pruning of MSA} 
\begin{figure}[h]
\includegraphics[width=0.40\textwidth]{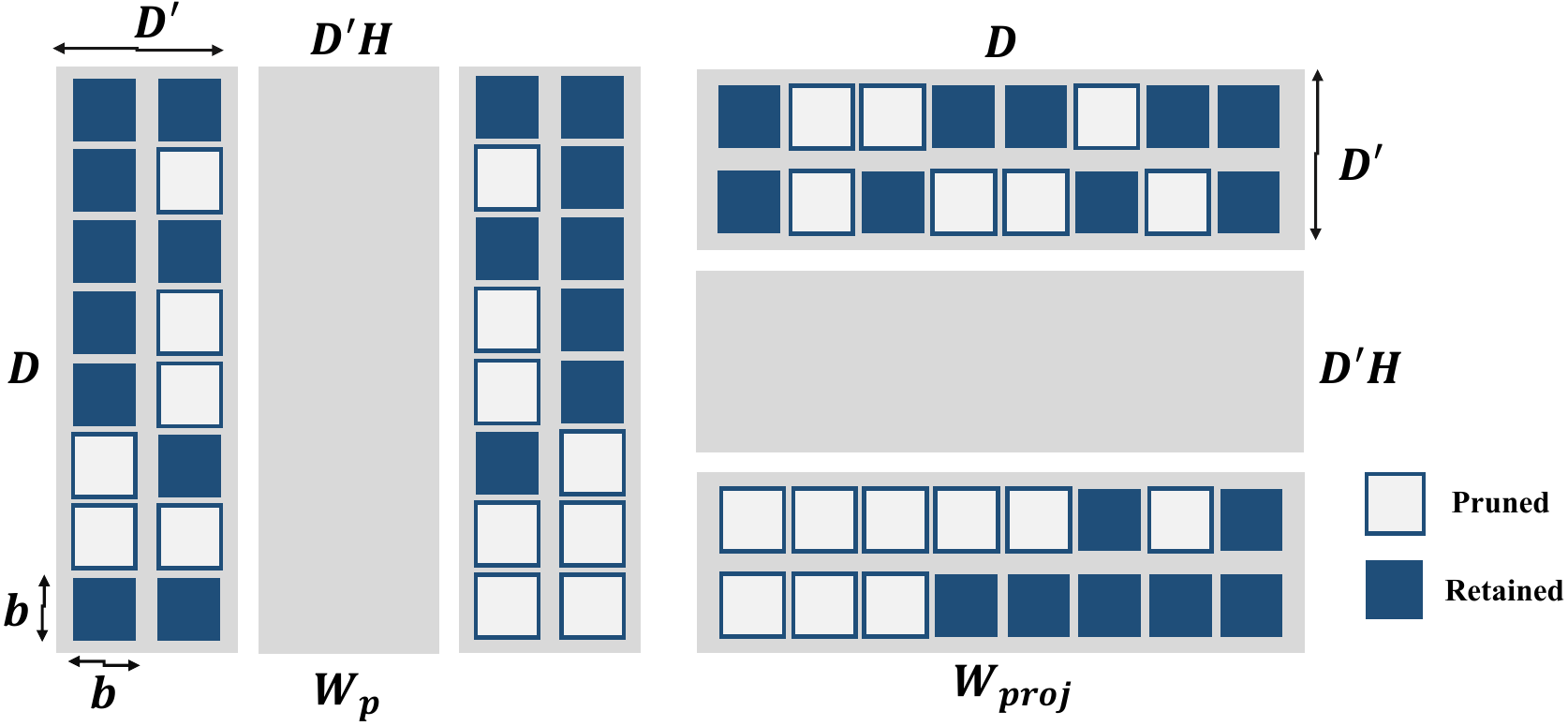}
\centering
\vspace{-0.3cm}
\caption{Alternate pattern of block pruning for $\mathbf{W}_p$ and $\mathbf{W}_{\text{proj}}$ parameters.}
\label{fig: msa_blk_pruning}
\vspace{-0.2cm}
\end{figure}

We use $\mathbf{W}_{q}$, $\mathbf{W}_{k}$, $\mathbf{W}_{v}$ $\in\mathbb{R}^{D \times HD'}$ to denote the concatenation of weight matrices of all the heads. For example, 
$
    \mathbf{W}_p \in \mathbb{R}^{D \times HD'} \text{ } \mbox{where} \text{ } p = \{q, k, v\}.
$
The projection operation (Equation \ref{eq: msa}) projects the concatenated SA outputs of embedding dimension $HD'$ back to dimension $D$ via 
$
    \mathbf{W}_{\text{proj}} \in \mathbb{R}^{HD' \times D}
$.
To prune a weight matrix $\mathbf{W} \in \mathbb{R}^{M_{1} \times M_{2}}$, we define a parameterized score matrix $\mathbf{S} \in \mathbb{R}^{m \times n}$ such that,
$
    (m, n)=(\ceil*{\frac{M_{1}}{b}}, \ceil*{\frac{M_{2}}{b}})
$
where $(b, b)$ is the block size. $\mathbf{S}_{ij}$ denotes the importance score of a parameter block of size $(b, b)$ in the weight matrix $\mathbf{W}$ denoted by the slice $\mathbf{W}(ib: \alpha,jb: \beta)$ where $(\alpha, \beta) = (\min(ib + b, M_{1}), \min(jb + b, M_{2}))$. 
% (slices exclude last element and indexing is from 0). 
$\mathbf{S}$ is used to construct a mask $\mathbf{M} \in \mathbb{R}^{M_{1} \times M_{2}}$ via the top-$k$ selection:

\begin{equation}
    \mathbf{M}_{ij}^{\text{block}}(s_{ij}) = 
    \begin{cases}
        1 & \text{if  } s_{ij} \in \text{top-$k$ of  } \mathbf{S} \\
        0 & \text{otherwise}
    \end{cases}
    \label{eq: mask_gen}
\end{equation}
wher $\mathbf{M}_{ij}^{\text{block}}(.)$ is a block of size $(b, b)$ in $\mathbf{M}$ corresponding to $s_{ij}$. 
% (redundant iunformation) Essentially, if a score value is in the top-K of score values in $\mathbf{S}$, we set the corresponding block in mask $\mathbf{M}$ to 1.  
The masked weight is generated as,
$
    \mathbf{W(\mathbf{M})} = \mathbf{W} \odot \mathbf{M}  
$
where $\odot$ is the element-wise Hadamard product.
The generated masked weight $\mathbf{W(\mathbf{M})}$ is used for the forward pass during training. Note that top-$k$ is the target weight blocks of interest. To compute the gradient of $\mathbf{S}$ during the backward pass, a straight-through estimator (STE) \cite{bengio2013estimating, ramanujan2020whats, mallya2018piggyback} is used that neglects the gradients of $\mathbf{M}$ with respect to $\mathbf{S}$.
% The block size $b$, for pruning parameters $\mathbf{W}_p$ and $\mathbf{W}_{proj}$, is set so that $D$ and $D'$ are both a multiple of $b$. This ensures structural uniformity when contiguous blocks in a head are pruned, leading to the removal of an entire attention head.
Additionally, the pruning of $\mathbf{W}_p$ in row dimension and the pruning of $\mathbf{W}_{\text{proj}}$ in column dimension follows the same pattern (denoted as \emph{alternate} pattern), as shown in Figure \ref{fig: msa_blk_pruning}. For example, a head removed from $\mathbf{W}_p$ makes the corresponding head in $\mathbf{W}_{\text{proj}}$ redundant, and vice-versa.

\begin{figure}[h]
\includegraphics[width=0.4\textwidth]{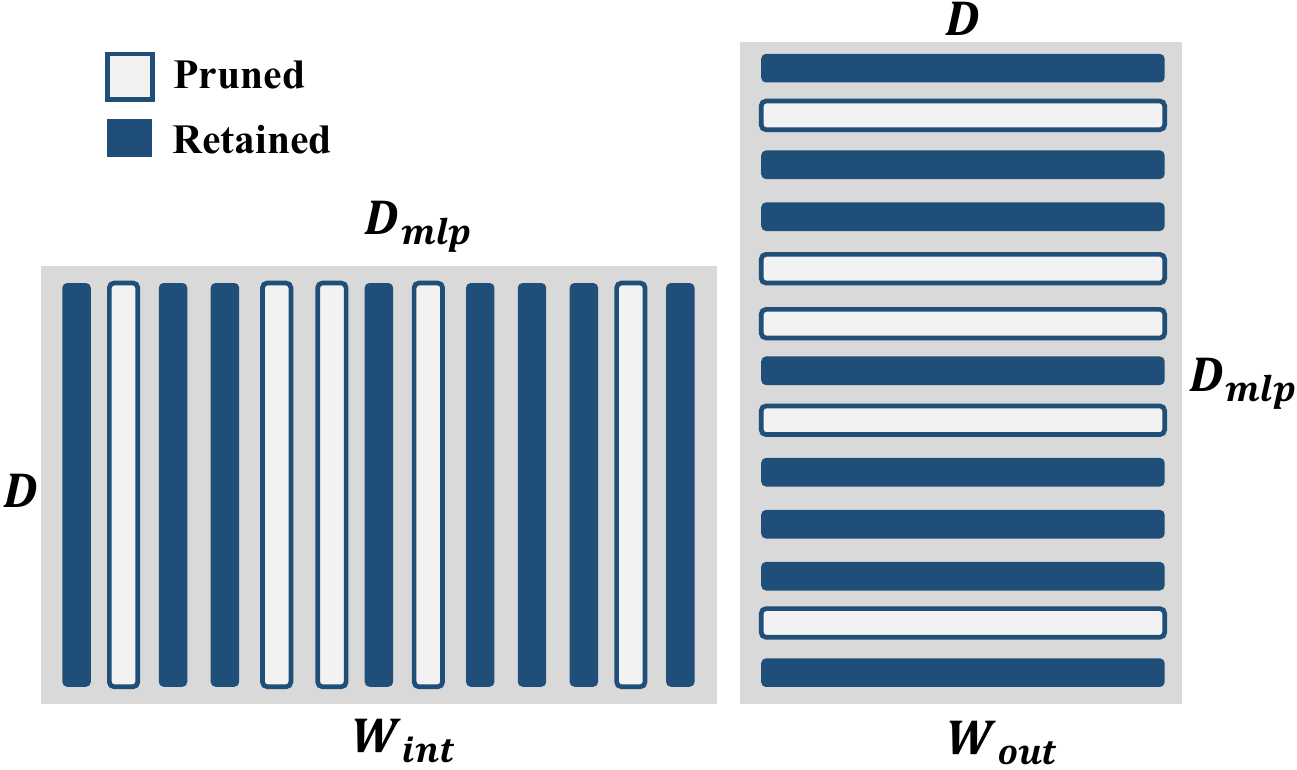}
\centering
\vspace{-0.3cm}
\caption{Alternate column-wise/row-wise pruning for $\mathbf{W}_{\text{int}}$ and $\mathbf{W}_{\text{out}}$. Note that $D_{mlp}$ is much larger than $D$. }
\label{fig: mlp_neur_prune}

\end{figure}

\textbf{MLP Pruning.} The weight matrices in MLP are
$
    \mathbf{W}_{\text{int}} \in \mathbb{R}^{D \times D_{\text{mlp}}} \quad \mbox{and} \quad \mathbf{W}_{\text{out}} \in \mathbb{R}^{D_{\text{mlp}} \times D} 
$.
The pruning of $\mathbf{W}_{\text{int}}$ and $\mathbf{W}_{\text{out}}$ follows the approach for pruning MSA. A key difference, however, is in how the score parameters are defined for  $\mathbf{W}_{\text{int}}$ and $\mathbf{W}_{\text{out}}$. Specifically, we define the scores as:
$
    \mathbf{S}_{\text{linear}} \in \mathbb{R}^{D_{\text{mlp}}} \text{ } \mbox{where} \text{ } \text{linear} = \{\text{int}, \text{out}\}
$.
The score vectors are defined to prune entire columns of $\mathbf{W}_{\text{int}}$ and  entire rows for $\mathbf{W}_{\text{out}}$ (see Figure \ref{fig: mlp_neur_prune}). 
% The former corresponds to removing entire neurons whilst the latter corresponds to removing parameters at fixed indices across all the neurons.  
Masks are generated column-wise/row-wise through top-$k$ selection. 
% As before, we prune $\mathbf{W}_{\text{int}}$ and $\mathbf{W}_{\text{out}}$ in an alternating fashion by removing rows of $\mathbf{W}_{\text{out}}$ corresponding to pruned columns in $\mathbf{W}_{\text{int}}$ and vice-versa (see Figure \ref{fig: mlp_neur_prune}). 
The natural parameter partitioning along the heads in MSA makes block pruning more effective in terms of removing entire heads. MLP parameters, on the other hand, lack such a partitioning. We thus focus on removing entire columns/rows for MLP parameters.
For model training, we add a norm of the sigmoid of scores to the training loss \cite{DBLP:journals/corr/abs-2109-04838_blk_prn_fstr_xmer}:
\begin{equation}
    \min_{\mathbf{W}} \mathcal{L} \rightarrow \min_{\mathbf{W}, \mathbf{S}} \mathcal{L} + \lambda||\sigma(\mathbf{S})|| \text{ } \mbox{where} \text{ } ||\mathbf{A}|| = \sum_{ij} A_{ij} 
    \label{eq: trn_lss}
\end{equation}
where the loss is updated to penalize the presence of a model parameter, thus driving the model to be sparse. The extent of this penalization is controlled by the hyper-parameter $\lambda$.

\subsection{Dynamic Token Pruning}
\label{subsec: dyn_tkn_prn}

\begin{figure}[h]
\includegraphics[width=0.43\textwidth]{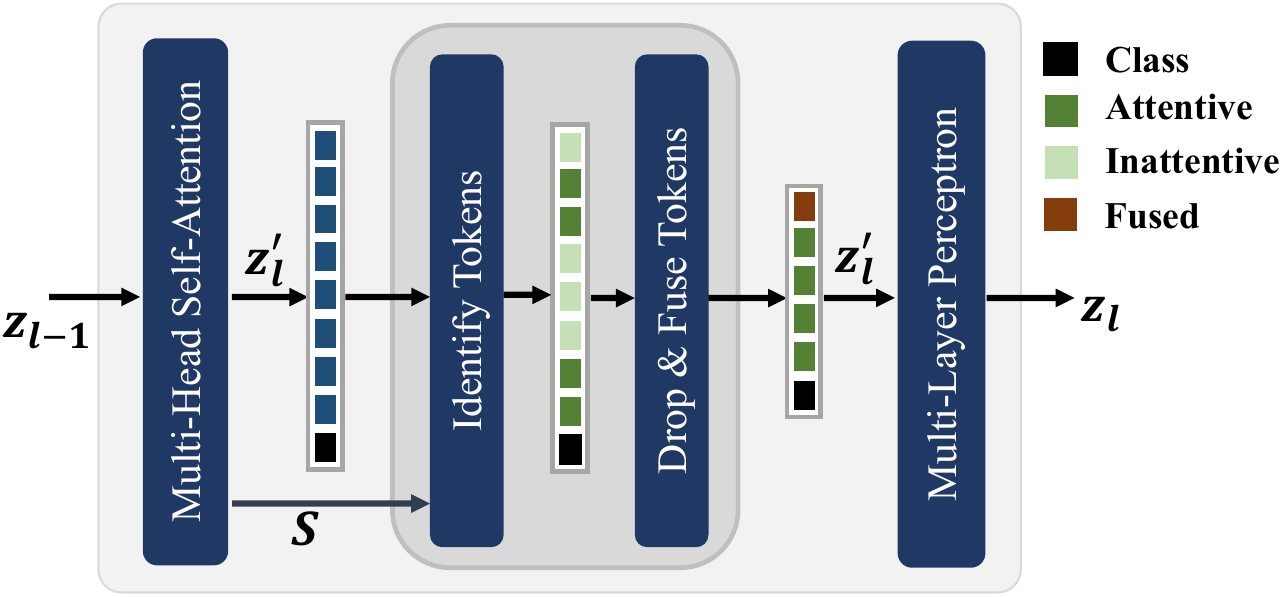}
\centering
\vspace{-0.3cm}
\caption{TDM inserted between the MSA and MLP block inside an encoder. TDM updates the input to the MLP block, $\mathbf{Z}_{l}{'}$, as $\mathbf{Z}_{l}{'} \leftarrow \mbox{TDM}(\mathbf{Z}_{l}{'})$.}
\label{fig: tokn_drop_modl}

\end{figure}

% Whilst  prior works focus on either pruning weights or on pruning the model inputs (section \ref{subsec: rel_work}), we propose a simultaneous approach to pruning that prunes model parameters as well as model inputs. 
Dynamic Token Pruning prunes along the token dimension $N$. The redundancy along the token dimension comes from the fact that several patches within an input image are inattentive, contributing insignificantly to the final learned model \cite{kong2022spvit, kim2022learned, liang2022patches, rao2021dynamicvit}. Since ViTs can inherently handle inputs with an arbitrary number of tokens (patches), we exploit this independence of the input token dimension from the model parameter dimension(s) by dropping inattentive tokens.
Specifically, to classify tokens into \emph{attentive} and \emph{inattentive} tokens, we use a non-parametric approach \cite{liang2022patches}. The attention $\mathbf{A}$ computed within the MSA (Equation \ref{eq: attention}) is utilized to perform attentive token identification. 
In MSA, the attention score $\mathbf{A}_h$ is generated by each head.
% Within a particular head $h$, with a corresponding attention score matrix $\mathbf{A}_h$, we focus on the attention scores of all tokens with respect to the class token $\mathbf{x}_{\text{CLS}}$ (excluding its attention score with itself) obtained by the slice $\mathbf{A}_h[0, 1:]$ (the class token is assumed to have a token index 0).
We aggregate the above score vector across all the heads using
$
    \mathbf{\mathcal{S}} = \frac{1}{H} \sum_{h} \mathbf{A}_h \text{ } \mbox{where} \text{ } \mathbf{\mathcal{S}} \in \mathbb{R}^{N}
$
and represents the importance score of every single token. Based on a keep-rate $r_t$, a total of $\ceil*{(N-1)r_t}$ tokens with the top scores in $\mathbf{\mathcal{S}}$ are retained. The remaining inattentive tokens are fused into a single token by performing a weighted aggregation of these tokens with respect to their respective scores in $\mathbf{\mathcal{S}}$.
The above token dropping is performed via a token dropping module (TDM) inserted between the MSA and the MLP modules (Figure \ref{fig: tokn_drop_modl}), with the tokens dropped dynamically during both training and inference.

\subsection{Simultaneous Pruning}
\label{subsec:simu-pruning}

To recover accuracy for the pruned model, we utilize the knowledge distillation technique commonly used to transfer knowledge from an already trained larger teacher model to a smaller student model \cite{chittyvenkata2023survey}. The class logits associated with the teacher and the student networks are used to compute a distillation loss at a distillation temperature $T$:
\begin{equation}
    \mathcal{L}_{\text{distill}} = T^2 \text{KL}(\mathbf{p}_{\text{teacher}}(T) \; || \; \mathbf{p_{\text{student}}} (T))
    \label{eq: dstil}
\end{equation}
where $\text{KL}(.)$ stands for the KL divergence loss. $\mathbf{p}(T)$ refers to the softmax probability vector computed at a temperature of $T$ from input logits vector $\mathbf{l}_p$ as $\frac{\text{exp}(\mathbf{l}_p/T)}{\sum_{i}exp(\mathbf{l}_p(i)/T)}$. The final loss is obtained as a weighted sum of the generic loss and the distillation loss, with the weights acting as hyper-parameters. The simultaneous training algorithm used to train a sparse model on sparse attentive input tokens is given in Algorithm \ref{alg: siml_fineprun}.   $\text{Encoder}_{\text{TDM}}^{\mathcal{M}^{s}, j}$ is an encoder at layer $j$ with the TDM module included, in a ViT model $\mathcal{M}^{s}$. Similarly, $\text{Encoder}^{\mathcal{M}^{s}, j}$ is an encoder at layer $j$ without the TDM.

\begin{algorithm}
\caption{Simultaneous Fine-Pruning}
\label{alg: siml_fineprun}
\begin{small}
\begin{algorithmic}[1]
    \Require Student model $\mathcal{M}^{s}(\mathbf{x}; \Theta)$; teacher model $\mathcal{M}^{t}(\mathbf{x})$; model pruning top-$k$ rate $r_b$; input token pruning keep rate $r_t$; set \{$\ell$\} of encoders at some depth in the model where TDM used; dataset $\mathcal{D}$ for fine-pruning
    \Ensure Set of weight and score parameters $\{\mathbf{W}, \mathbf{S}\}$ are initialized
    
    \For{$i=1...\text{epochs}$}
        \For{all $\mathbf{x}$ in $\mathcal{D}$}
            \State{Compute masks \{$\mathbf{M}$\} using scores \{$\mathbf{S}$\} via $r_b$} 
            \State{$\mathbf{W}(\mathbf{M}) \gets \mathbf{W} \odot \mathbf{M}$ for all $\mathbf{W} \in \{\mathbf{W}\}$}
            \State{$\mathbf{y} \gets \mathbf{x}$}
            \For {encoders in $\mathcal{M}^{s}$ at  layer $j$ from $1...L$}
                \If {$j \in \{\ell\}$}
                    \State{$\mathbf{y} \gets \text{Encoder}_{\text{TDM}}^{\mathcal{M}^{s}, j}(\mathbf{y})$}
                \Else 
                    \State{$\mathbf{y} \gets \text{Encoder}^{\mathcal{M}^{s}, j}(\mathbf{y})$}
                \EndIf
            \EndFor
        \State{Compute student logits $\mathbf{z}_s$ from $\mathbf{y}$ and final classifier of $\mathcal{M}^{s}$} 
        \State{Compute teacher logits $\mathbf{z}_t$ using $\mathcal{M}^{t}(\mathbf{x})$ and final classifier of $\mathcal{M}^{t}$}
        \State{Compute $\mathcal{L}_{distill}$ via $\mathbf{z}_t$, $\mathbf{z}_s$ and Euquation \ref{eq: dstil}}
        \State{Compute $\mathcal{L}$ as in Euquation \ref{eq: trn_lss}}
        \State{$\mathcal{L}_{\text{net}} \gets \lambda_{\text{distill}}\mathcal{L}_{\text{distill}} + \lambda_{\text{normal}}\mathcal{L}$}
        \State{Backpropogate $\mathcal{L}_{\text{net}}$} and compute gradients
        \State{Update $\{\mathbf{W}, \mathbf{S}\}$}
        \EndFor
    \EndFor
\end{algorithmic}
\end{small}
\end{algorithm}

\subsection{Computational Complexity: Pruned Model}

We analyze the computational complexity for the proposed pruned model. The complexity of an encoder is described in table \ref{tab:complexity_pruned}. $\alpha$ is the average ratio of retained weight blocks to the total  weight blocks (retained and pruned) within a column of blocks in parameter matrices $\mathbf{W}_p$ (computed after the removal of heads pruned in their entirety). 
$\alpha'$ is defined similarly, but for matrix $\mathbf{W}_{\text{proj}}$. $H_{\text{kept}}$ are the number of heads retained within MSA. 
$N_{\text{kept}}$ are the total retained tokens after token dropping ($\approx Nr_t$). $\alpha^{\text{mlp}}$ is the ratio of retained neurons (same for both $\mathbf{W}_{\text{int}}$ and $\mathbf{W}_{\text{out}}$). Note that $\alpha^{\text{mlp}} = r_b$.

\begin{table}[h]
\vspace{-0.2cm}
    \centering
    \caption{Computational Complexity of Pruned Model}
    \vspace{-0.3cm}
    \begin{adjustbox}{max width=0.37\textwidth}
    \begin{tabular}{cc}
         \toprule
         \textbf{Operation} & \textbf{Computational Complexity} \\
         \midrule
         \midrule
         LayerNorm 1 ($\times 1$) & $BND$ \\
         \midrule
         LayerNorm 2 ($\times 1$) & $BN_{\text{kept}}D$ \\
         \midrule
         Residual Add 1 ($\times 1$) & $BND$ \\
         \midrule
         Residual Add 2 ($\times 1$) & $BN_{\text{kept}}D$ \\
         \midrule
         MSA ($\times 1$) & $BH_{\text{kept}}ND'D(3\alpha + \alpha')+2BH_{\text{kept}}N^{2}D'$ \\
         \midrule
         TDM ($\times 1$) & $BN(H+N+D)$ \\
         \midrule
         MLP ($\times 1$) & $2BN_{\text{kept}}DD_{\text{mlp}}\alpha^{\text{mlp}}$ \\
         \midrule
         \begin{tabular}[|c|]{@{}c@{}} \textbf{Total} \\\textbf{Complexity} \end{tabular} & \begin{tabular}[|c|]{@{}c@{}} $2BND + 2BN_{\text{kept}}D + BH_{\text{kept}}ND'D(3\alpha + \alpha')$ \\ $+ 2BH_{\text{kept}}N^{2}D' + BN(H+N+D)$ \\ $ + 2BN_{\text{kept}}DD_{\text{mlp}}\alpha^{\text{mlp}}$ \end{tabular} \\
         \bottomrule
    \end{tabular}
    \end{adjustbox}
    
    \label{tab:complexity_pruned}
\end{table}

%Few details more left

\begin{figure}[]
\includegraphics[width=0.4\textwidth]{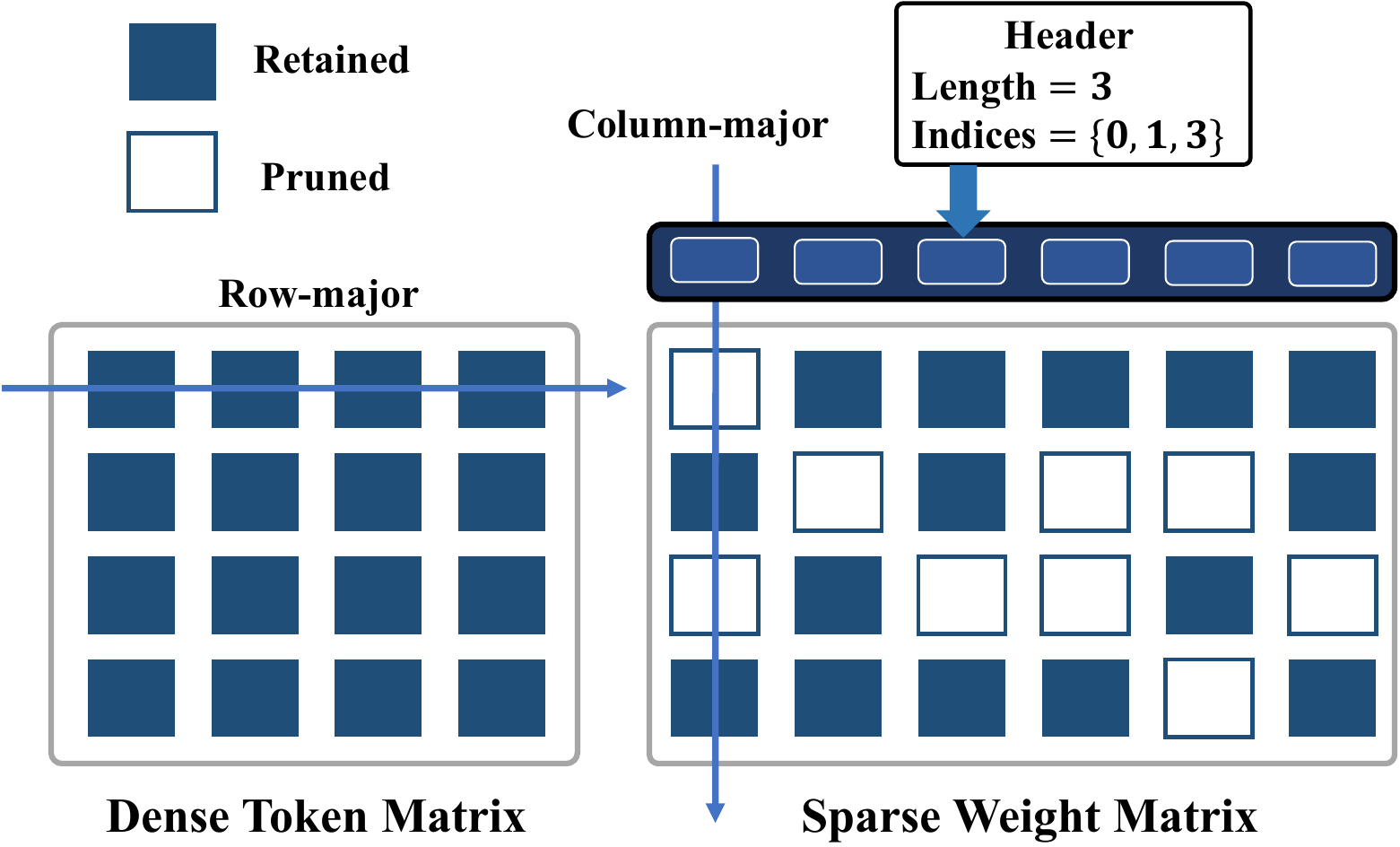}
\centering
\vspace{-0.3cm}
\caption{Data layout of dense token matrix and sparse weight matrix.}
\vspace{-0.4cm}
\label{fig: data_layout}
\end{figure}

\section{Hardware Design}
\label{sec:Hardware-Design}

In this Section, we introduce our hardware design to accelerate the pruned model on the FPGA platform. To be specific: in Section \ref{subsec: hw_data_org}, we introduce the data format and layout that store the sparse (and dense) weight matrices and input data; in Section \ref{subsec: hw_ovw}, we introduce the main components in the proposed hardware architecture. In Section \ref{subsec: hw_wrkflw}, we introduce the workflow for executing the pruned ViT encoder using the proposed hardware design.

 \begin{figure*}[h]
\includegraphics[height=0.35\textwidth]{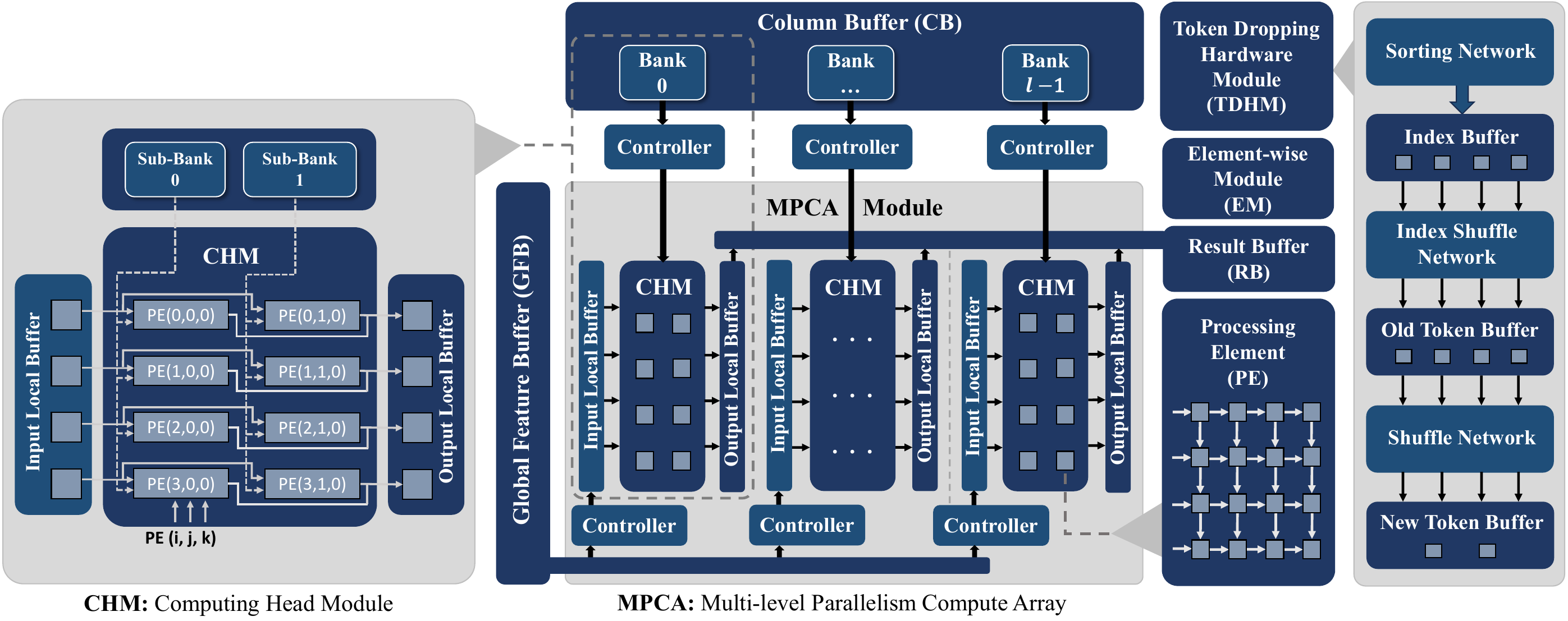}
\centering
\vspace{-0.4cm}
\caption{Overview of hardware architecture.}
\vspace{-0.4cm}
\label{fig: hw_overvw}
\end{figure*}

 \begin{figure}[h]
\includegraphics[width=0.48\textwidth]{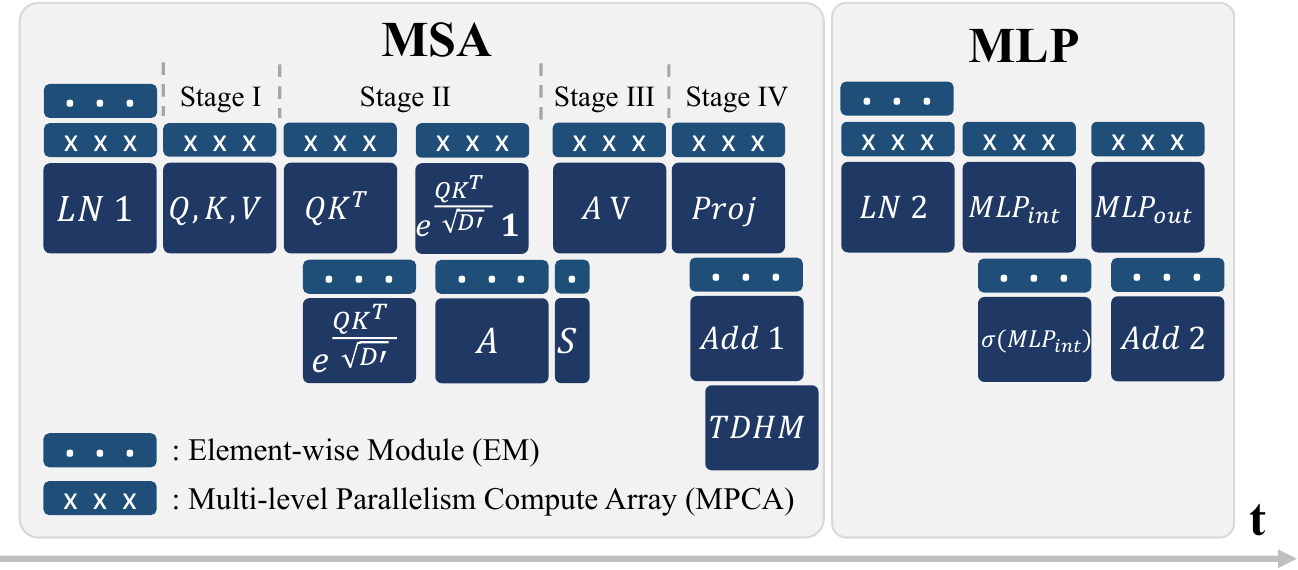}
\centering
\vspace{-0.3cm}
\caption{Task scheduling for executing an ViT encoder on the proposed architecture. $LN$ denotes LayerNorm}
\vspace{-0.4cm}
\label{fig: compute_timeline}
\end{figure}

% Add subsection and flow of subsections later

\subsection{Data Format and Layout}
\label{subsec: hw_data_org}

Due to structured block pruning, all weight and feature matrices are partitioned into data \emph{blocks} of the same size $b\times b$. All the data in the same blocks are stored contiguously. The dense matrix is stored in block-wise \emph{row-major} order such that all the data blocks of the same row are stored contiguously in memory space.
The weight matrices are stored in \emph{column-major} order such that all the unpruned data blocks in the same column are stored contiguous in memory space as shown in Figure \ref{fig: data_layout}. Note that for the sparse weight matrices, only unpruned data blocks are stored.
% (trival thing, can be removed) Bias are packed together with their corresponding weight matrices. 
For each column in a block-wise sparse weight matrix, we include a \emph{header} at its beginning that encodes row indices of the present blocks and the length of the column block. For simplicity, in the rest of the paper, a row of matrix denotes a row of data blocks, and a column of matrix denotes a column of data blocks.

%

%For sparse weight matrices, we only store the dense parameter blocks (unpruned blocks), in column-major order.  

\subsection{Hardware Overview}
\label{subsec: hw_ovw}

As shown in Figure \ref{fig: hw_overvw}, the architecture design comprises of: (1) Multi-level Parallelism Compute Array (MPCA), (2)  Element-wise Module (EM), (3) Token Dropping Hardware Module (TDHM). Besides, there are on-chip buffers, including a Global Feature Buffer (GFB) that stores the feature matrices, (2) Column Buffers (CB) that store the weight matrix, (3) Result Buffers (RB) that store the results of the current layer.

In MPCA, the computation units are organized into multiple levels. An MPCA has $p_{\text{h}}$ parallel Computing Head Modules (CHMs). Each CHM has a 2-D array of Processing Elements (PEs) of size $p_{\text{t}} \times p_{\text{c}}$. Each Processing Element has an array of computation units of size $p_{\text{pe}} \times p_{\text{pe}}$. Essentially, $p_{\text{h}}$, $p_{\text{t}}$, $p_{\text{c}}$, $p_{\text{pe}}^{2}$ are the computation parallelism in the head dimension, input token dimension, weight column dimension, and the data parallelism within the data blocks, respectively. The Element-wise Module (EM) performs element-wise GELU and exponentiation. The Token Dropping Hardware Module (TDHM) performs dynamical token dropping (Section \ref{subsec: dyn_tkn_prn}).  

The computation units in MPCA are organized to multi-level because (1) it enables massive data parallelism in MSA and MLP,
% To be specific, the  $p_{\text{h}}$ parallel CHMs can compute multiple heads of MSA in parallel; the parallelism of $p_{\text{t}}$ enables the CHM to compute multiple row blocks in the input token matrices in parallel; the parallelism on $p_{\text{c}}$ enables the CHM to compute multiple column blocks in weight matrices; $p_{\text{pe}}^{2}$ denotes the data parallelism within each data block ($b\times b$), 
(2) it enables data reuse/sharing within CHM. The PEs in the same column of CHM can share the same weight block, while the PEs in the same rows of CHM can share the same input token block. This data-sharing strategy simplifies the computation complicated by the irregular data access pattern of block-wise weight pruning.  (3) By selecting proper $p_{\text{c}}$ with a load balancing strategy, we can alleviate the load imbalance caused by block-wise weight pruning. 

%%%%%%%%%%%%%%%%%%%%%%%%%%%%%%%%%%%% NOTE %%%%%%%%%%%%%%%%%%%%%%%%%%%%%%%%%%%%%%%%%%%%%%%%%%%%%%%%
%%%%%%%%%%% Edit portions pertaining to local input buffer so that its correct/clear %%%%%%%%%%%%%
%%%%%%%%%%%%%%%%%%%%%%%%%%%%%%%%%%%%%%%%%%%%%%%%%%%%%%%%%%%%%%%%%%%%%%%%%%%%%%%%%%%%%%%%%%%%%%%%%%
% -------------------------------- Very Important Clarification ----------------------------------
% Local Input Buffer don't contain whole "copies" but just see whole "copies" (and their versions)
% This needs to be clarified
% Minimally, a single line clarification; can be explained in the performance model section
%%%%%%%%%%%%%%%%%%%%%%%%%%%%%%%%%%%%%%%%%%%%%%%%%%%%%%%%%%%%%%%%%%%%%%%%%%%%%%%%%%%%%%%%%%%%%%%%%%

\subsection{Workflow}
\label{subsec: hw_wrkflw}

%\noindent \textbf{Executing the MSA}: The computation of a MSA involves $H$ heads (See Equation \ref{eq: msa}). \textcolor{red}{[to Dhurv]: describe how the MPCA executes the MSA, The weight matrices and token matrices will be partitioned into blocks of size $b\times b$ ($b$ is the size for weight parition). After partitioning, you can describe how to execute the MSA.}
%\noindent \textbf{Executing the MLP}:  \textcolor{red}{[to Dhurv]: describe how the MPCA executes the MLP}
%\noindent \textbf{On-the-fly Token Dropping}: \textcolor{red}{[to Dhurv]: describe how the token dropping hardwae module executes the token dropping}

% The proposed accelerator executes the input model layer-by-layer with each layer executed using the same set of computational resources (e.g., MPCA). Each layer as follows:

The proposed accelerator executes the input model layer-by-layer, with each layer being executed using the same set of computational resources (e.g., MPCA). Each CHM utilizes Sparse Block-wise Matrix Multiplication (SBMM) to process sparse matrices in blocks. For dense matrices, the computation shifts to Dense Block-wise Matrix Multiplication (DBMM) and Dense Head-wise Block Matrix Multiplication (DHBMM) for a focused computation on matrix blocks associated with each head. Each layer performs as follows:

\vspace{0.1cm}
\subsubsection{MSA Execution} 
\begin{figure}[h]
\includegraphics[width=0.42\textwidth]{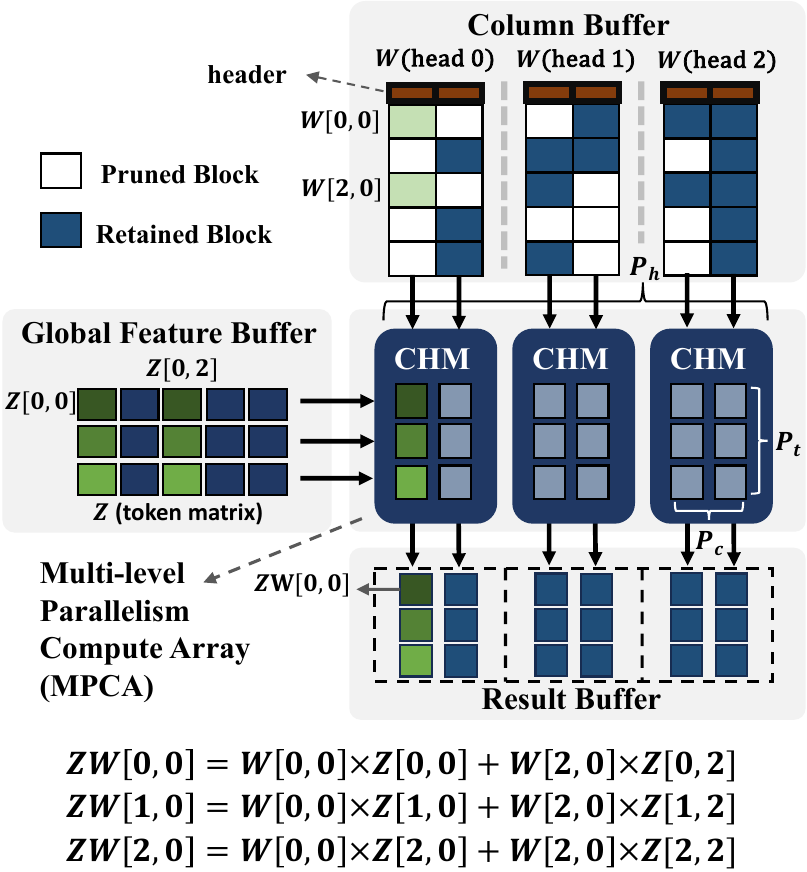}
\centering
\vspace{-0.1cm}
\caption{Execution of  Sparse Block-wise Matrix Multiplication (SBMM) on MSA. Note that $X[i,j]$ denotes a data block at $i^{\text{th}}$ row and $j^{\text{th}}$ column. In Each CHM, the PEs of the same column share the same column block of weight matrix. In Each CHM,  the PEs of the same row block of token matrix.}
\label{fig: SBMM_kernel}
\end{figure}

MSA is divided into four stages (shown in Figure \ref{fig: compute_timeline}):  stage (i) computes $\mathbf{Q}$, $\mathbf{K}$ and $\mathbf{V}$ through $[\mathbf{Q}, \mathbf{K}, \mathbf{V}] = \mathbf{Z}[\mathbf{W}_q,  \mathbf{W}_k, \mathbf{W}_v]$. $\mathbf{Z}$ is the input token matrix and $[\;]$ is matrix concatenation. 
% Note that $\mathbf{Q}, \mathbf{K}$ and $\mathbf{V}$ refer to the query, key, and value matrices of all the heads concatenated.
Let $\mathbf{Q}_{h}, \mathbf{K}_{h}$ and $\mathbf{V}_{h}$ denote the query, key, and value matrices for a head $h$, where $(0\leq h < H)$. % This compute corresponds to the Sparse Block-wise Matrix Multiplication (SBMM) primitive, executed via the MPCA module as shown in Figure \ref{fig: SBMM_kernel}. 
The algorithm for executing each matrix multiplication (e.g., dense token matrix $\mathbf{Z}$ multiply by sparse weight matrices $\mathbf{W}_q,  \mathbf{W}_k, \mathbf{W}_v$) in MSA using MPCA is shown in Algorithm \ref{alg: SBMM/DBMM} and an example is shown in Figure \ref{fig: SBMM_kernel}.
Each CHM  computes its corresponding head $[\mathbf{Q}_{h}, \mathbf{K}_{h}, \mathbf{V}_{h}]$. Since we perform block partitioning for each matrix, the computation is executed in a block-wise fashion. 
Within each CHM, the PEs within a column share the same column of weight, which is stored in the column buffer (CB). 
PEs of the same row share the same row of $\mathbf{Z}$.
We use $\text{PE}(i, j, k)$ to denote a PE in the $k^{\text{th}}$ CHM at location $(i,j)$ within the CHM. A column of PEs: $\text{PE}(:, j, k)$ share the same column of weight with the corresponding header information (See Figure \ref{fig: data_layout}). 
Thus, each PE in $j^{\text{th}}$ column utilizes the shared header indices  (Figure \ref{fig: data_layout}) to fetch the corresponding data block from the token matrix (in the local input buffer) to perform block-wise matrix multiplication.
% using only the feature row blocks corresponding to the non-zero weight blocks.
% associated with the PE column to slice the local CHM input buffer along the corresponding block row $i$ before feeding the blocks to the $\text{PE}(i,j,k)$ 
The partial results are accumulated in local result buffers.
% before being written out to the global result buffer and finally the external memory. 

The stage (ii) computes the attention scores $\mathbf{A}_{h}=\mbox{softmax}(\frac{\mathbf{Q}_{h} \mathbf{K_{h}}^{T}}{\sqrt{D'}}), \text{ } (0 \leq i < H)$. $\mathbf{Q}_{h}\mathbf{K}_{h}^{T}$ is dense block-wise matrix multiplication executed via the MPCA module (See Algorithm \ref{alg: SBMM/DBMM}). $\mathbf{Q}$ is buffered in the GFB, and $\mathbf{K}^{T}$ is buffered in the CB. 
% A key difference from stage (i) is that since the DBMM is performed head-wise, instead of entire row blocks being broadcasted from the GFB to the local input buffers for each CHM, 
% We broadcast row blocks within a head of $\mathbf{Q}$ ($\mathbf{Q}_{h}$) to its corresponding CHM. 
The output data blocks of $\mathbf{Q}_{h}\mathbf{K}_{h}^{T}$ are sent to EM module for element-wise scaling (by $1/\sqrt{D'}$) and exponentiation to obtain $\text{exp}(\frac{\mathbf{Q}_{h}\mathbf{K}_{h}^{T}}{\sqrt{D'}})$. Then, we utilize MPCA to compute the scaling factors for $\mbox{softmax}(\frac{\mathbf{Q}_{h}\mathbf{K}_{h}^{T}}{\sqrt{D'}})$.
% by performing a head-wise DBMM with row blocks of $\text{exp}(\frac{\mathbf{Q}_{h}\mathbf{K}_{h}^{T}}{\sqrt{D'}})$, for each head, being placed in the local input buffer of its corresponding CHM. Each bank of the CB contains a single column vector $\mathbf{1} \in \mathbb{R}^{N}$, with each entry $1$, and partitioned into blocks, each block of shape $(b, 1)$. With this setting, MPCA aggregates each row of the matrix $\text{exp}(\frac{\mathbf{Q}_{h}\mathbf{K}_{h}^{T}}{\sqrt{D'}})$, for all $h$. Note that for this compute with the MPCA, only the first column of PEs in each CHM will be active (with only the first column of compute units active within each PE).
The rows of matrix $\text{exp}(\frac{\mathbf{Q}_{h}\mathbf{K}_{h}^{T}}{\sqrt{D'}})$ and their corresponding computed scaling factors, are streamed from MPCA to EM to perform the scaling to obtain the attention scores, $\mathbf{A}^{H}$.

The stage (iii) computes the self-attention $\mathbf{A}_{h}\mathbf{V}_{h}$. It is similar to the computation of $\mathbf{Q}_{h}\mathbf{K}_{h}^{T}$. The stage (iv) computes the projection (Equation \ref{eq: msa}).
% $[\mathbf{A}_{0}\mathbf{V}_{0}, \mathbf{A}_{1}\mathbf{V}_{1}, ..., \mathbf{A}_{H-1}\mathbf{V}_{H-1}]\mathbf{W}_{\text{proj}}$.
It is similar to the computation of $\mathbf{Q}$, $\mathbf{K}$ and $\mathbf{V}$ described in stage (i) as $\mathbf{W}_{\text{proj}}$ is the block-wise sparse matrix due to pruning.

\subsubsection{MLP Execution}

Since the weights of MLP are pruned for entire columns or rows (for $\mathbf{W}_{\text{int}}$ and $\mathbf{W}_{\text{out}}$ respectively), MLP layers can be mapped into dense block-wise matrix-matrix multiplication executed by MPCA (Algorithm \ref{alg: SBMM/DBMM}). This computation is similar to the computation of MSA (computing $\mathbf{Q}\mathbf{K}^{T}$). 
% Since $D_{\text{mlp}}$ is generally much larger than $D$ and $D = HD'$, we can partition the compute into multiple heads computed in parallel by each CHM as when computing $\mathbf{Q}$, $\mathbf{K}$ and $\mathbf{V}$. 
GELU activation is computed using the EM module.

\begin{algorithm}
\caption{Executing Sparse Block-wise Matrix Multiplication (SBMM) and Dense Block-wise Matrix Multiplication  (DBMM) through multi-level parallelism of MPCA}

\label{alg: SBMM/DBMM}
\begin{small}
\begin{algorithmic}[1]
    \Require Input matrix $\mathbf{X} \in \mathbb{R}^{M_{1} \times M_{2}}$; weight matrix $\mathbf{W}=[\mathbf{W}_{0},\mathbf{W}_{1},...,\mathbf{W}_{H-1}] \in \mathbb{R}^{M_{2} \times D}$, where each $\mathbf{W}_{h}\in \mathbb{R}^{M_{2} \times D'}$ ($0\leq h < H$); $D=HD'$ where $H$ denotes the number of heads and $D'$ is the dimension per head; block size $b$ % notation should be consistent with prior Section
    % (we have introduced it before, do not need to redefine here) $p_{\text{h}}$ CHM units of size $p_{\text{t}} \times p_{\text{c}}$ each such that $\text{CHM}_k$ represents the $k^{\text{th}}$ such CHM and $\text{PE}_{k}(i, j)$ represents the $(i, j)^{\text{th}}$ PE within $\text{CHM}_{k}$ with $p_{\text{pe}} \times p_{\text{pe}}$ compute units within the PE itself 
    
    \Ensure Output matrix $\mathbf{Y} = \mathbf{X}\mathbf{W} \in \mathbb{R}^{M_{1} \times D}$ 
    
    \State // $\mathbf{X}$ and $\mathbf{Y}$ are stored in block-wise row-major order and $\mathbf{W}$ is stored in block-wise column-major order (Figure \ref{fig: data_layout})
    
    \State // $\mathbf{X}[i,j]$ denotes the $(i,j)^{\text{th}}$ block of size $b \times b$ in $\mathbf{X}$ 
    % such that $i = 0 \; \text{to} \; \ceil*{\frac{M}{b}} - 1$ and $j = 0 \; \text{to} \; \ceil*{\frac{N}{b}} - 1$ 
    
    \State // $\mathbf{W}_{h}[i,j]$ denotes the $(i, j)^{\text{th}}$ block of size $b \times b$ in  $\mathbf{W}_{h}$ 
    % where $h = 0 \; \text{to} \; H-1$, $i = 0 \; \text{to} \; \ceil*{\frac{N}{b}}-1$, $j = 0 \; \text{to} \; \ceil*{\frac{q}{b}}-1$. $\mathbf{W}_{h} = \mathbf{W}[:,hD':(h+1)D']$

    \State // $\mathbf{Y}_{h}$ is the output corresponding to $\mathbf{W}_{h}$
    
    \State // {\color{blue} To compute $H$ heads, $p_{h}$ CHMs need $\ceil*{\frac{H}{p_{\text{h}}}}$ iterations}
    \For{$i = 0\; \text{ to }\;\ceil*{\frac{H}{p_{\text{h}}}}-1$} %\footnote{}  {\color{blue}\Comment{See footnote}}
        \For{each $\text{CHM}_j$ with $j = 0\;\text{ to }\;p_{\text{h}}-1$ \textbf{Parallel}}
            \State // {\color{blue} $\text{CHM}_j$ computes $\mathbf{Y}_{j+ip_{\text{h}}}$}
            \State // {\color{blue} To compute $\ceil*{\frac{D'}{b}}$ column blocks of a $\mathbf{W}_{h}$, $p_{c}$ columns \hspace*{3.2em} of PEs in a CHM need $\ceil*{\ceil*{\frac{D'}{b}}/p_{\text{t}}}$ iterations}
            \For{$k = 0 \;\text{ to }\; \ceil*{\ceil*{\frac{D'}{b}}/{p_{\text{c}}}} - 1$} 
            %\footnote{}   {\color{blue}\Comment{See footnote}}
            \State // {\color{blue}Load weights into CB}
                \State // {\color{blue} To compute $\ceil*{\frac{M}{b}}$ row blocks of $\mathbf{X}$, $p_{t}$ rows of PEs \hspace*{5em} in a CHM need $\ceil*{\ceil*{\frac{M}{b}}/p_{\text{t}}}$ iterations.}
                \For{$l = 0 \;\text{ to }\; \ceil*{\ceil*{\frac{M}{b}}/p_{\text{t}}}-1$}  
                %\footnote{} {\color{blue}\Comment{See footnote}}
                \State // {\color{blue}Load data (partition of $\mathbf{X}$) into GFB}
                    \For{each $\text{PE}_{j}(m, n)$ in $\text{CHM}_{j}$ \textbf{Parallel}}
                        \State // {\color{blue}$\text{PE}_{j}(m,n)$ computes output block \hspace*{9em} $\mathbf{Y}_{j + ip_{\text{h}}}[m + lp_{\text{t}},n+kp_{\text{c}}]$}
                        \If{MPCA mode is SBMM}
                            \State Fetch $\mathbf{X}[m+lp_{\text{t}}, \text{idx}]$ from GFB for all idx \hspace*{10em}  in the header of $\mathbf{W}_{j+ip_{\text{h}}}[:, n+kp_{\text{c}}]$
                            \State Compute $\mathbf{Y}_{j + ip_{\text{h}}}[m + lp_{\text{t}},n+kp_{\text{c}}]$ using \hspace*{10em}  the fetched input blocks from GFB 
                        \Else
                            \State // {\color{blue}MPCA mode is DBMM}
                            \State Fetch $\mathbf{X}[m+lp_{\text{t}}, :]$ from GFB
                            \State Compute $\mathbf{Y}_{j + ip_{\text{h}}}[m + lp_{\text{t}},n+kp_{\text{c}}]$ using \hspace*{10em}  all the input blocks from GFB 
                        \EndIf
                        
                    \EndFor
                \EndFor
            \EndFor
        \EndFor
    \EndFor
\end{algorithmic}
\end{small}
\end{algorithm}

\begin{comment}
\footnotetext{[\textbf{Line 5 of Algorithm \ref{alg: SBMM/DBMM}}] To compute $H$ heads, $p_{h}$ CHMs need $\ceil*{\frac{H}{p_{\text{h}}}}$ iterations}
\footnotetext{[\textbf{Line 8 of Algorithm \ref{alg: SBMM/DBMM}}] To compute $\ceil*{\frac{D'}{b}}$ column blocks of a $\mathbf{W}_{h}$, $p_{c}$ columns of PEs in a CHM need $\ceil*{\ceil*{\frac{D'}{b}}/p_{\text{t}}}$ iterations}
\footnotetext{[\textbf{Line 10 of Algorithm \ref{alg: SBMM/DBMM}}] To compute $\ceil*{\frac{M}{b}}$ row blocks of $\mathbf{X}$, $p_{t}$ rows of PEs in a CHM need $\ceil*{\ceil*{\frac{M}{b}}/p_{\text{t}}}$ iterations.}
\end{comment}

%%%%% Traverse this paragraphs meaning to paragraphs above 
%%%%% shorten/remove by clarifying top portions

% Note that the local input buffers for a CHM, at a given time, only contain the current feature row blocks required to perform a single block-wise matrix multiplication for all the PEs. For DBMM, this is shared across PEs within a row, whilst for SBMM, PEs in a row can get different row blocks at a given time based on the sparsity pattern of the column blocks they are associated with. Broadcasting, with respect to the local input buffers, refers here to the entire set of row blocks that PEs within a CHM see, over the course of computing their respective output blocks.

%% DERRICKS FIGURE MUCH BETTER, SO USED INSTEAD of mpca_compute.pdf
%\begin{figure}[H]
%\includegraphics[width=0.40\textwidth]{figures/mpca_compute.pdf}
%\centering
%\caption{ }
%\label{fig: SBMM_kernel}
%\end{figure}

\subsubsection{Dynamic Token Dropping}
\label{subsubsec:Dynamic-Token-Dropping}

We design a token dropping hardware module (TDHM) for  on-the-fly  token dropping  and reorganizing the remaining tokens. 
% The savings associated with the reduction of tokens, for a ViT, are thus unblemished by the additional computation required to select the attentive tokens. 
% In particular, if an encoder contains a TDM (Figure \ref{fig: tokn_drop_modl}), the TDHM starts its execution as soon as the attention scores are available during the MSA execution.
The token pruning is based on importance scores of tokens $\mathbf{\mathcal{S}}$ (See Secton \ref{subsec: dyn_tkn_prn}). The attention scores $\mathbf{A}_h$ for all the heads are buffered in the TDHM as soon as they are computed via MSA execution. Then, scores $\mathbf{\mathcal{S}} = \frac{1}{H} \sum_{h} \mathbf{A}_h$ are computed via the EM.
% (after stage (ii) of MSA excution, once all the attention scores $\mathbf{A}^{H}$ are computed).
After that, a bitonic sorting network sorts the scores $\mathbf{\mathcal{S}}$ to obtain the indices of top-$k$ tokens. Original, each token has a row index in the input token matrix $\mathbf{Z}_{\text{in}}$, denoted as \emph{old token index} ($\text{id}_{\text{old}}$). After sorting by   $\mathbf{\mathcal{S}}$, each token is assigned \emph{new token index} ($\text{id}_{\text{new}}$)  which is the row index in the output token matrix $\mathbf{Z}_{\text{out}}$. Therefore, the sorting network generates ($\text{id}_{\text{old}}$, $\text{id}_{\text{new}}$, $\text{flag}$) for each token, where $\text{flag}$ indicates if the token will be pruned. To organize output token matrix $\mathbf{Z}_{\text{out}}$ (stored in Old Token Buffer),
an index shuffle network routes each $(\text{id}_{\text{old}}, \text{id}_{\text{new}}, $\text{flag}$)$ to Old Token Buffer for fetching tokens according to $\text{id}_{\text{old}}$.
Then, the fetched tokens are routed to the New Token Buffer according to $\text{id}_{\text{new}}$, which generates too-$k$ token matrix and non-top-$k$ token matrix.
The non-top-$k$ tokens are then fused into a single token and merged with the top-$k$ tokens to produce the output of TDHM module.

\subsection{Optimizations}
\label{subsec: opt}

\subsubsection{Load balancing across columns for SBMM} 
As the weight matrices are pruned in block-wise fashion, different columns in a weight matrix can have different number of data blocks, which can potentially lead to load imbalance.
% Sparse computations are typically bottlenecked by the load imbalance that occurs across PEs that compute on parameters varying in sparsity levels. 
The multi-level parallelism in MPCA distributes the PEs across several heads (CHMs).
PEs within each CHM computes on weight column blocks using multiple iterations that each iteration executes different columns of weight matrices (See Algorithm \ref{alg: SBMM/DBMM}).
This naturally reduces the impact of load imbalance due to differing sparsity levels across the columns of weight matrices. 
As we restrict block-wise pruning to only the weights of MSA ($\mathbf{W}_{q}$, $\mathbf{W}_{k}$, $\mathbf{W}_{v}$, $\mathbf{W}_{\text{proj}}$), this further reduces the impact of load imbalance.
Note that the gain in performance by block-wise pruning the MSA parameters (leading to the removal of entire heads) is far outweighted by the load imbalance presented by such pruning. 
Prior methods \cite{9424344} that balance such block-wise pruning across columns are disadvantaged by the fact that they cannot remove entire heads. 
Moreover, we perform offline workload assignment among columns of weight marices, prior to inference, such that workloads of columns are evenly distributed across different columns of PEs within a CHM.

%%%%%%%%%%%%%%%%%%%%%%%%%%%%%%%%%%%%%%%%%%%%%%%%%%%
%%%%%%% ADD that density up-scaled by H_orig/H_kept * density which improves load balancing %%%%%%%%%%%%%%

\subsubsection{Dealing with varying number of tokens, retained heads and block sizes} Dynamic token pruning, leads to varying number of tokens for different layers. Moreover, block-wise weight pruning the MSA parameters leads to removal of heads within an encoder. In general, the heads removed or retained in each encoder can vary, which can potentially lead to runtime hardware underutilization. For example, if the number of rows of token matrix $\frac{N}{b} < p_{t}$, $p_{t} - \frac{N}{b}$ rows of PEs in a CHM will be idle. 
As we utilize multi-level parallelism in MPCA, through selecting proper $p_{\text{t}}$ (parallelism in token dimension) and $p_{\text{h}}$ (parallelism in head dimension), we can alleviate the underutilization. We use $\frac{N_{\text{min}}}{b}$ to denote the minimum number of row blocks of all the intermediate token matrices and use $H_{\text{min}}$ to denote the  minimum number of heads of all the layers. Through setting $p_{\text{t}} \ll \frac{N_{\text{min}}}{b}$, the PEs utilization in a CHM will be $>\frac{\frac{N_{\text{min}}}{p_{t }\times b}  }{\lceil \frac{N_{\text{min}}}{p_{t }\times b} \rceil}$ (Suppose $6\times p_{\text{t}} < \frac{N_{\text{min}}}{b}$. The utilization will be $>85\%$ ). Similar, we can set $p_{\text{h}} \ll H_{\text{min}}$.

% The MPCA dimensions $p_{\text{t}}$ and $p_{\text{h}}$ are closely related to the token blocks and the retained heads, respectively.

% their variation across the network requires a careful selection of the above parameters to ensure maximum resource utilization whilst satisfying the resource constraints. To this end, we utilize an offline profiler that sweeps over a set of feasible hardware parameters to select the optimal parameters subject to the total resources allocated, utilization of allocated resources and the model inference performance (latency). This also supports varying block sizes for block-pruning with the block size $b$ being related to the size of the PE compute unit $p_{\text{pe}}$.

% (as we decide not to put DSA in the paper, we can remove this part) 

% \subsubsection{Matching between the MPCA, EM and TDHM units} To allow for maximum concurrency between the compute units via an efficient streaming dataflow, we match the PE mesh between the MPCA and the EM units. Further, we also match the buffer sizes used across all three compute units - MPCA, EM and TDHM.

% \subsubsection{Buffer size reduction and weight reuse} Pruning model weights and input tokens simultaneously leads to a reduction in the buffer sizes required for the compute units. The SBMM and DBMM computation (Algorithm \ref{alg: SBMM/DBMM}) is optimized for weight reuse. Particularly, the PEs in each CHM of the MPCA compute along entire output column blocks before moving horizontally to compute the next set of output column blocks. 

\subsection{Resource and Performance Models}

%\textcolor{red}{[Message]: write a paragraph to define the hyperparameters of the hardware: $P_{\text{PE}}$, $P_{\text{Token}}$, $P_{\text{Column}}$, $P_{\text{head}}$}

%\textcolor{red}{[Message]: write a paragraph and equations to explain the number of cycles for executing the MSA, $P_{\text{PE}}$, $P_{\text{Token}}$, $P_{\text{Column}}$, $P_{\text{head}}$}

%\textcolor{red}{[Message]: write a paragraph and equations to explain the number of cycles for executing the MLP, $P_{\text{PE}}$, $P_{\text{Token}}$, $P_{\text{Column}}$, $P_{\text{head}}$}

\subsubsection{Resource Consumption Model} We perform theoretical analysis for the performance achieved by the codesign and its hardware resource utilization. We denote the total computational resources utilized by the MPCA, EM and TDHM as $R_{\text{MPCA}}, R_{\text{EM}}$ and $R_{\text{TDHM}}$, respectively. $R_{\text{MPCA}}$ is proportional to the total number of computation units: $p_{\text{t}}p_{\text{h}}p_{\text{c}}p_{\text{pe}^2}$. Compared to $R_{\text{MPCA}}$, the resources used by $R_{\text{TDHM}}$ and $R_{\text{EM}}$ are negligible, and thus ignored for analysis. The total $R_{\text{Total}}$ (DSPs and LUTs) are 
%\begin{equation}
 $R_{\text{Total}} = (c_{1}p_{\text{t}}p_{\text{h}}p_{\text{c}}p_{\text{pe}}^{2}, 
    c_{2}p_{\text{t}}p_{\text{h}}p_{\text{c}}p_{\text{pe}}^{2})$,
%\label{eq: net_res}
%\end{equation}
where $c_{1}$ and $c_{2}$ denote the amount of DSPs and LUTs utilized by a single computation unit. The size of the (global) feature buffer, column buffer and the (global) result buffer, associated with the MPCA, are $b^{2}p_{t}\gamma$, $b^{2}p_{c}\gamma$ and $b^{2}p_{\text{t}}p_{\text{h}}p_{\text{c}}$, respectively. Here, $b$ is the block size and $\gamma$ is a constant that equals the (maximum) total number of row blocks required to compute a single output block. We match the buffer sizes across compute units to improve the dataflow performance of the accelerator. This gives a total buffer size of $4\times \max({b^{2}p_{\text{t}}p_{\text{h}}p_{\text{c}}, b^{2}p_{t}\gamma})$ for the EM module  and $2 \times \max({b^{2}p_{\text{t}}p_{\text{h}}p_{\text{c}}, b^{2}p_{t}\gamma})$ for the TDHM module. The EM module requires a buffer to store the input, scaling factor, addtion factor and the output. Similarly, TDHM requires an input and an output buffer. The total size of required buffers $B_{\text{Total}}$ is given as,
%\begin{equation}
%\begin{split}
    $B_{\text{Total}} = b^{2}p_{t}\gamma + b^{2}p_{c}\gamma + b^{2}p_{\text{t}}p_{\text{h}}p_{\text{c}} + 6\times \max({b^{2}p_{\text{t}}p_{\text{h}}p_{\text{c}}, b^{2}p_{t}\gamma})$.
%\end{split}
%\end{equation}
$R_{\text{Total}}$ and $B_{\text{Total}}$ are the estimation of resource utilization. 
The main design $p_{\text{t}}$, $p_{\text{h}}$ and $p_{\text{c}}$, with $\gamma$ are empirically set according the resource of target FPGA platform (See Section \ref{sec: impl_det} for details).
% The parameters $p_{\text{t}}$ and $p_{\text{h}}$ are heuristically selected based on the number of token row blocks $\ceil*{\frac{N}{b}}$ and the number of retained heads $H_{\text{kept}}$ appearing across encoders in the network, respectively, by trading-off compute concurrency with resource utilization. $p_{\text{c}}$ is generally restricted to be $2$ to allow for concurrent memory access (see \ref{sec: impl_det}) and $p_{\text{pe}}$ is set such that the block size $b$ is a multiple of $p_{\text{pe}}$.

\subsubsection{Performance Model} Based on algorithm $\ref{alg: SBMM/DBMM}$, the number of cycles to perform either SBMM, DBMM or DHBMM is estimated in table \ref{tab: cycle_count}. Note that DHBMM is DBMM computed head-wise (as in stage (ii) of MSA execution). In table \ref{tab: cycle_count}, the cycles for SBMM/DBMM are the cycles required to multiply a matrix of dimension $(M_{1}, M_{2})$ with a matrix of dimension $(M_{2}, D)$. $D'$ is the size per head, $b$ is the block size and $\phi$ is the ratio of retained dense blocks to total blocks within a column of the matrix. Note that for DBMM, $\phi$ is $1$ and for SBMM, $\phi$ is assumed similar in each column block for simplicity.
$(M_{1}, M_{2})$ and $(M_{2}, D)$ are the per head left and right matrix sizes, for DHBMM, with $H$ being the total number of heads. The cycle estimates in Table \ref{tab: cycle_count} can be used to compute the total cycles for the MSA and the MLP blocks.

\begin{table}[]
\centering
\caption{Execution cycles for SBMM/DBMM and DHBMM}
\vspace{-0.3cm}
\label{tab: cycle_count}
\begin{tabular}{cc}
\toprule
& \textbf{Cycles} \\
\midrule
\textbf{SBMM/DBMM} & $\ceil*{\frac{\ceil*{\frac{M_{1}}{b}} \ceil*{\frac{D'}{b}}}{p_{\text{t}}p_{\text{c}}}} \ceil*{\frac{\ceil*{\frac{D}{D'}}}{p_{\text{h}}}} \ceil*{\frac{M_{2}}{b}} \ceil*{\frac{b}{p_{\text{pe}}}}^{2} b \phi $ \\
\textbf{DHBMM} & $\ceil*{\frac{\ceil*{\frac{M_{1}}{b}} \ceil*{\frac{D}{b}}}{{p_{\text{t}}p_{\text{c}}}}} \ceil*{\frac{H}{p_{\text{h}}}} \ceil*{\frac{M_{2}}{b}} \ceil*{\frac{b}{p_{\text{pe}}}}^{2} b  $      \\
\bottomrule
\end{tabular}
\vspace{-0.3cm}
\end{table}
\section{Implementation Details}
\label{sec: impl_det}

% Add subsection for Design Space Exploration?

% \begin{table}[h!]
% \centering
% \caption{Model hyperparamters of DeiT-Small}
% \label{tab:model-parameters}
% \begin{adjustbox}{max width=0.48\textwidth}
% \begin{tabular}{|cc|cc|} 
% \hline  \textbf{Total Parameters} & 22 million  &  \textbf{Layers} & 12 \\ \hline
% \textbf{Attention Heads} & 6  & \textbf{Hidden Dimension ($D$)} & 384 \\  \hline
% \textbf{MLP Ratio} & 4  & \textbf{Image Size} & $224 \times 224$ \\  \hline
% \textbf{Patch Size} & $ 16 \times 16$ &  &\\ 
% \hline
% \end{tabular}
% \end{adjustbox}
% \end{table}

\noindent \textbf{Evaluated Model}: We evaluate our approach on the widely used DeiT-Small \cite{touvron2021trainingdeit} model, which has 12 layers, with each layer having six heads. The hidden dimension is $D=384$, and the (base) model has $22$M parameters.

\vspace{0.1cm}
\noindent \textbf{Implementation details of weight pruning, token pruning, and simultaneous training}: The DeiT-Small model is simultaneously pruned as per algorithm \ref{alg: siml_fineprun}. We train several variants of the model by varying the model pruning top-$k$ rate $r_{b}$, token pruning keep rate $r_{t}$, and block size $b$. Specifically, $r_{b}$ is varied over $\{0.5, 0.7\}$, $r_{t}$ over $\{0.5, 0.7, 0.9\}$ and $b$ over $\{16, 32\}$. A cubic sparsity scheduler, as in \cite{sanh2020movement}, is used to schedule $r_{b}$ from a full density of $1$ to its final density ($0.5$ or $0.7$) with a warm-up and a cool-down phase. The token-dropping module, TDM, is inserted in the $3^{\text{rd}}$, $7^{\text{th}}$ and $10^{\text{th}}$ encoder layers. All model variants are trained for a total of $30$ epochs using the AdamW optimizer \cite{loshchilov2019decoupled} with a learning rate of $2\times10^{-5}$, weight decay of $0.01$ and a batch size of $128$ across $4$ GPUs. Note that to train all pruned variants as well as the baseline, we use the pre-trained DeiT-Small model available at \cite{hgnfcmdl} with the classification MLP head parameters re-initialized. A ViT base model is used as the teacher for knowledge distillation.

\vspace{0.1cm}
\noindent  \textbf{Hardware implementation details}: We implement our FPGA design on a state-of-the-art FPGA platform, Xilinx Alveo U250, which consists of four Super Logic Regions (SLRs). We implement the proposed hardware design using Xilinx High-level Synthesis (HLS). For the MPCA module, we empirically determine the hardware hyperparameters to be $p_{h}=4$, $p_{t} = 12$, $p_{c} = 2$, $p_{pe} = 8$ according to the hardware resources of the target FPGA board:
(1) We set $p_{h}=4$ because Alveo u250 has four SLRs, with each SLR placed in one CHM.
(2) We set $p_{c}=2$ because in a CHM, the PEs of the same row load the same rows of tokens but different data blocks. The BRAM/URAM on FPGA has two independent memory ports, which can support concurrent memory access of $2$ columns of PEs ($p_{c}=2$).
(3) We set $p_{pe} = 8$ because the data block size $b$ is set as $16$ or $32$ for block-wise weight pruning. Using $p_{pe} = 8$ can support these two block sizes without data padding as well as keeping a reasonable value for $p_{t}$.
(4) For setting $p_{t}$, the PEs of the same column within a CHM shares the same weight blocks. The weight blocks are broadcast into each PE of the same column, which supports any value of $p_{t}$. We set $p_{t}$ according to the available resources of the target FPGA board after determining $p_{h}$, $p_{c}$, and $p_{pe}$. We use the int16 data format. We utilize the four DDR4 channels of U250, which have 77 GB/s of external memory bandwidth in total. We perform synthesis and place-route for the design using Xilinx Vitis v2022.2.  We report the frequency and FPGA resource utilization after place-route. The achieved frequency is $300$ MHz, and the resource utilization is shown in Table \ref{tab:fpga-resource}.

\begin{table}[h]
\centering
\vspace{-0.3cm}
\caption{FPGA resource utilization}
\vspace{-0.3cm}
\label{tab:fpga-resource}
\begin{adjustbox}{max width=0.4\textwidth}
\begin{tabular}{ccccc}
\toprule
 & \textbf{LUTs} &  \textbf{DSPs} & \textbf{URAMs} & \textbf{BRAMs}  \\
 \midrule
\vspace{0.1cm}
\textbf{HeatViT} \cite{dong2023heatvit} & 137.6K\-161.4K & 1955-2066 & N/A &  338-528 \\
\vspace{0.1cm}
\textbf{Auto-Vit-Acc} \cite{li2022autovitacc} & 120K-193K & 13-2066 & N/A & N/A \\
\textbf{Our Work} & 798K & 7088 & 1728 & 960  \\
 \bottomrule
\end{tabular}
\end{adjustbox}
\vspace{-0.3cm}
\end{table}

\begin{comment}
\textbf{0.5-0.75 pages} 

\textcolor{red}{[Message]: write a paragraph to explain implementation details on the FPGA board, and spec of the FPGA board}

\textcolor{red}{[Message]: write a paragraph to explain how we select the hardware hyperparameters $P_{\text{PE}}$, $P_{\text{Token}}$, $P_{\text{Column}}$, $P_{\text{head}}$}
\end{comment}
\section{Experiments and Results}

\label{sec: exp_res}

\subsection{Baselines, Metrics, Datasets}
%\textbf{Baselines work (CPU, GPU, ViT FPGA works) to compare} Check Table V of Dynasparse paper

\noindent \textbf{Baselines:} We compare our implementation on FPGA with the state-of-art CPU, GPU, and FPGA accelerators including \cite{dong2023heatvit} and \cite{wang2022vitaccelerator}. Table \ref{tab:platform-specs} shows the details of these platforms.

\begin{table}[h!]
\centering
\caption{Specifications of platforms}
\vspace{-0.3cm}
\label{tab:platform-specs}
\begin{adjustbox}{max width=0.48\textwidth}
\begin{tabular}{cccccc} 
\toprule
& \textbf{CPU} & \textbf{GPU} & \textbf{HeatViT} \cite{dong2023heatvit} & \textbf{SPViT} \cite{kong2022spvit} & \textbf{Our work} \\ 
\midrule
\midrule
\textbf{Platform} &  \begin{tabular}[|c|]{@{}c@{}} AMD \\ EPYC 9654 \end{tabular} & \begin{tabular}[|c|]{@{}c@{}}  NVIDIA RTX \\ 6000 Ada \end{tabular}  & \begin{tabular}[|c|]{@{}c@{}}  Xilinx \\ ZCU102 \end{tabular} & \begin{tabular}[|c|]{@{}c@{}} Xilinx \\ ZCU102  \end{tabular} & \begin{tabular}[|c|]{@{}c@{}} Xilinx \\ Alveo U250 \end{tabular} \\  \midrule
\textbf{Frequency} & 2.4 GHz & 915 MHz & 150 MHz &  200 MHz& 300 MHz \\ \midrule  
\begin{tabular}[|c|]{@{}c@{}} \textbf{Peak} \\ \textbf{Performance} \\  \textbf{(TFLOPS)} \end{tabular}& 3.69 & 91.06 & 0.37 &  0.54 & 1.8 \\ \midrule  
\begin{tabular}[|c|]{@{}c@{}}  \textbf{On-chip} \\ \textbf{Memory}  \end{tabular} & 384 MB & 96MB & 3.6MB  & 4MB & 36 MB \\ \midrule
\begin{tabular}[|c|]{@{}c@{}} \textbf{Memory} \\ \textbf{Bandwidth}  \end{tabular}  & 461 GB/s & 960 GB/s & 19.2 GB/s &  19.2 GB/s & 77 GB/s\\
\bottomrule
\end{tabular}
\end{adjustbox}
\vspace{-0.35cm}
\end{table}

\noindent  \textbf{Datasets:} Following prior works \cite{liang2022patches}\cite{dong2023heatvit}, we use ImageNet dataset in our experiments with approximately 1.2 million images to evaluate our approach. 

% It is a known benchmark in computer vision providing an challenging collection of images for classification tasks. We want to remain consistent with other token pruning work like \cite{liang2022patches}\cite{dong2023heatvit} in terms of dataset usage. By using this dataset, we can ensure that our findings are relevant to most ViT works and our results are comparable with established outcomes.

\noindent  \textbf{Performance Metrics}:
We utilize the following performance metrics: (1) \emph{Accuracy:} Following prior works, we evaluate the accuracy of our pruned model on ImageNet. (2) \emph{Inference latency:} Following prior works \cite{dong2023heatvit, wang2022vitaccelerator, kong2022spvit}, we measured inference latency via hardware emulation using AMD-Xilinx Vitis, which accurately simulates the behavior of FPGA DDR. The measured latency is end-to-end from the time when the input is given at DDR to the time when the inference result is written back to DDR. (3) \emph{Throughput:} Throughput denotes the number of images that can be processed for a given time frame. (4) \emph{Computation complexity (FLOPS):} We measure the computational complexity, which is the number of floating-point operations (FLOPs). (5) \emph{Model size:} The amount of memory space (MB) to store the model.

\begin{table*}[]
\centering
\caption{The experimental results for different pruning settings}
\label{tab: results_fpga}
\vspace{-0.3cm}
\begin{adjustbox}{max width=0.98\textwidth}
\begin{tabular}{c|cc|c|c|cc|c|c|c|c}
\toprule
\multirow{2}{*}{\textbf{Notion}}& \multicolumn{2}{c|}{\textbf{Block Pruning}}  &\textbf{Token Pruning}  & \multirow{2}{*}{ \begin{tabular}[|c|]{@{}c@{}} \textbf{Head} \\ \textbf{Retained Ratio} \end{tabular} } & \multicolumn{2}{c|}{\textbf{Model Parameters}}   & \multirow{2}{*}{\begin{tabular}[|c|]{@{}c@{}} \textbf{Training} \\ \textbf{Epochs} \end{tabular}} & \multirow{2}{*}{\textbf{Accuracy}} & \multirow{2}{*}{ \begin{tabular}[|c|]{@{}c@{}} \textbf{FPGA} \\ \textbf{Latency} \\  (ms) \end{tabular} } & \multirow{2}{*}{ \begin{tabular}[|c|]{@{}c@{}} \textbf{FPGA} \\  \textbf{Throughput} \\ (images/second)\\\end{tabular} } \\ \cmidrule{2-4} \cmidrule{6-7}
& Block Size $b$ & Top-$k$ Rate $r_{b}$ & Token Keep Rate  $r_{t}$ &  & Model Size & MACs &  &  &  &  \\ \midrule
(Baseline) & 16 & 1 & 1 & 1 & 22M & 4.27G & 30 & 79.59\% & 3.19 & 313.00 \\
(Baseline) & 32 & 1 & 1 & 1 & 22M & 4.27G & 30 & 79.59\% & 3.55 & 281.43 \\

(Pruned)   & 16 & 0.5 & 0.5 & 0.91 & 14.29M & 1.32G & 30 & 66.86\% & 0.868 & 1151.55 \\
(Pruned)   & 16 & 0.5 & 0.7 & 0.91 & 14.29M & 1.79G & 30 & 68.62\% & 1.169 & 855.12 \\
(Pruned)   & 16 & 0.5 & 0.9 & 0.93 & 14.39M & 2.43G & 30 & 70.14\% & 1.479 & 676.10 \\

(Pruned)   & 16 & 0.7 & 0.5 & 0.98 & 17.63M & 1.62G & 30 & 74.12\% & 1.140 & 877.054 \\
(Pruned)   & 16 & 0.7 & 0.7 & 0.98 & 17.63M & 2.20G & 30 & 75.96\% & 1.553 & 643.72 \\
(Pruned)   & 16 & 0.7 & 0.9 & 0.98 & 17.63M & 2.98G & 30 & 76.55\% & 1.953 & 511.94 \\

(Pruned)   & 32 & 0.5 & 0.5 & 0.84 & 13.80M & 1.25G & 30 & 67.25\% & 1.621 & 616.79 \\
(Pruned)   & 32 & 0.5 & 0.7 & 0.83 & 13.70M & 1.70G & 30 & 68.62\% & 1.796 & 556.66\\
(Pruned)   & 32 & 0.5 & 0.9 & 0.84 & 13.80M & 2.31G & 30 & 70.06\% & 1.999 & 500.17\\

(Pruned)   & 32 & 0.7 & 0.5 & 0.97 & 17.53M & 1.61G & 30 & 73.45\% & 2.126 & 470.33\\
(Pruned)   & 32 & 0.7 & 0.7 & 0.94 & 17.33M & 2.16G & 30 & 75.65\% & 2.353 & 424.93\\
(Pruned)   & 32 & 0.7 & 0.9 & 0.94 & 17.33M & 2.93G & 30 & 76.40\% & 2.590 & 386.02 \\
\bottomrule
\end{tabular}
\end{adjustbox}
\vspace{-0.35cm}
\end{table*}

\subsection{Evaluation for the Pruning Algorithm}

Results in Table \ref{tab: results_fpga} indicate that for extreme pruning settings ($r_{b}$, $r_{t}$ both $0.5$), the accuracy drop ($\approx 12\%$) compared against the baseline is not insignificant. A major reason for this drop is the fact that the training epochs for our experiments were restricted to $30$ despite the reduction in model and input density. With a lower top-$k$ rate $r_{b}$ and token keep rate $r_{t}$, the model requires larger epochs to converge. Compared to the baseline DeiT-Small model, the proposed simultaneous pruning algorithm achieves a compression ratio of up to $1.24\times$ to $1.60\times$ and a reduction in the computational cost of up to $1.43\times$ to $3.42\times$ with an accuracy drop of as little as $\approx 3\%$. Whilst prior works focus on either reducing the model size \cite{9424344} or on reducing the computational complexity \cite{dong2023heatvit, kong2022spvit}, our proposed simultaneous pruning algorithm targets both.    

%\textbf{Impact of block pruning}

%\textbf{under different block sizes} 16x16, 32x32

%explain why not use 8x8 64x64

%\textbf{Different sparsity (weight matrices) level} 0.2 0.5

%\textbf{Different token pruning ratio} 0.9 0.5

%2x2x2 = 8

%\textbf{Obtain accuracy, FLOPs, inference latency/throughput on the CPU and GPU, }

%\textbf{Discussion}: talk about how to increase the accuracy further

%\textbf{accuracy, pruning ratio, FLOPS for Deit-Small (from pruning algorithm paper)}

% \begin{table}[]
% \centering
% \caption{Model Accuracy Comparing with other Token Pruning Algorithms}
% \label{tab:pruning-baselines}
% \begin{tabular}{cccc}
% \toprule
% \textbf{Keep Ratio} & \textbf{0.9} &  \textbf{0.7} & \textbf{0.5} \\
%  \midrule
% \textbf{Token Reorg\cite{liang2022patches}} &  &  &  \\ [1ex]
% \textbf{SPViT\cite{kong2022spvit}} &  &  &  \\[1ex]
% \textbf{Our work} &  &  &   \\
%  \bottomrule
% \end{tabular}
% \end{table}

\subsection{Evaluation on the FPGA accelerator}

\begin{figure}[h]
\vspace{-0.42cm}
\includegraphics[width=0.47\textwidth]{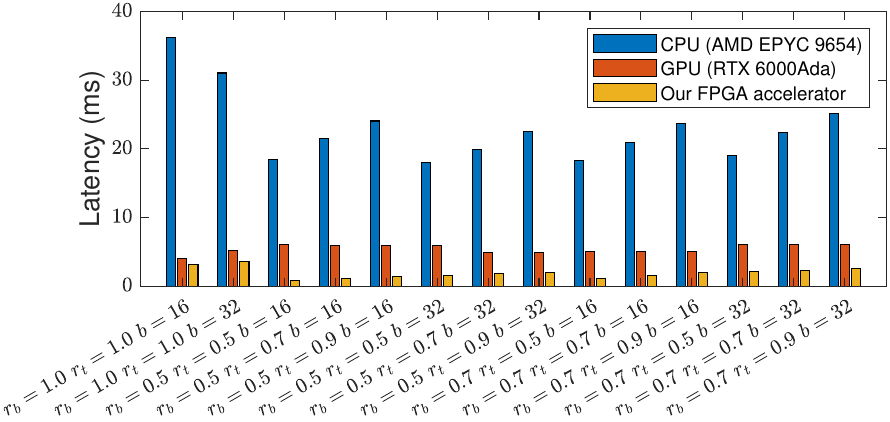}
\centering
\vspace{-0.35cm}
\caption{Comparison of latency under various pruning settings when batch size is 1 for all platforms. CPU, GPU, and FPGA execute the same model.}
\label{fig:cmp-latency}
\vspace{-0.3cm}
\end{figure}

\begin{figure}[h]
\includegraphics[width=0.47\textwidth]{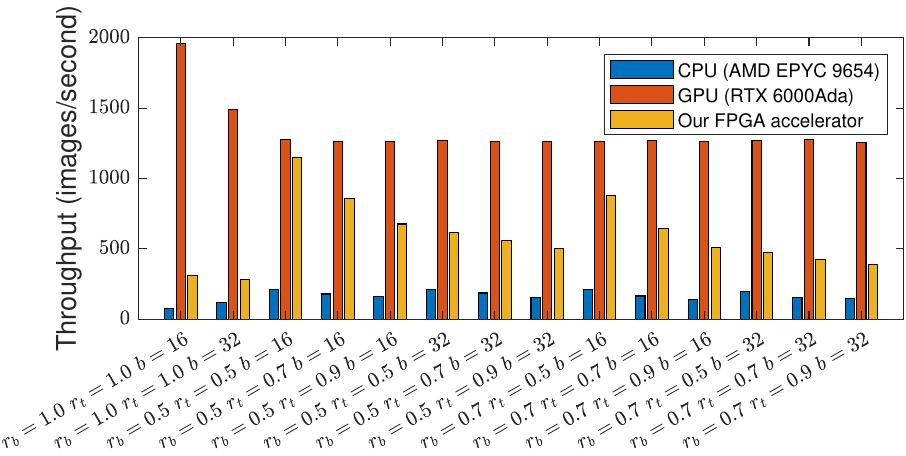}
\centering
\vspace{-0.3cm}
\caption{Comparison of throughput under various pruning settings when batch size is 8 for CPU and GPU and batch size is 1 for FPGA.}
\label{fig:cmp-throughput}
\vspace{-0.3cm}
\end{figure}

\subsubsection{Cross platform comparison}
We compare the latency and throughput for executing the pruned model with baseline CPU and GPU (Figure \ref{fig:cmp-latency} and \ref{fig:cmp-throughput}). The latency of our accelerator is measured when batch size is $1$, and the throughput is calculated by $\frac{1}{\text{latency}}$. For comparing the latency, we set the batch size as $1$ for CPU and GPU because a larger batch size will increase the latency for CPU and GPU. For throughput comparison, we set the batch size as $8$ for CPU and GPU, which can fully exploit their thread-level parallelism. On average, our FPGA accelerator achieves a latency reduction of $12.8\times$ and $3.2\times$, compared with CPU and GPU, respectively. The lower latency of our FPGA accelerator is due to the followings: (1) our MPCA module with load balance strategy fully exploits the computation parallelism within the pruned model. The FPGA accelerator achieves a higher speedup with higher pruning ratios (smaller $r_{b}$ and $r_{t}$). In contrast, the CPU and GPU cannot efficiently handle the computational irregularity caused by weight pruning and dynamic token dropping. (2)  CPU and GPU have complex cache hierarchies, leading to higher memory access latency for executing ViT inference, leading to increased latency.
On average, our FPGA accelerator achieves $3.6\times$ and $0.45\times$ throughput speedup compared with CPU and GPU, respectively. Our FPGA accelerator achieves a lower throughput ($0.45\times$) than GPU, because GPU has much higher peak performance ($50\times$) and eternal memory bandwidth. When the pruning ratios become high (e.g., $r_{b}=0.5$ and $r_{t}=0.5$), our throughput gets closer to GPU, which indicates that our FPGA accelerator has higher efficiency for executing the ViT model with larger pruning ratio.

\subsubsection{Comparison with state-of-the-art}
We compare the proposed codesign with the state-of-the-art ViT Accelerators \cite{li2022autovitacc, dong2023heatvit, kong2022spvit} on FPGA as shown in Table \ref{tab:hardware-baselines}. Prior works use at most one pruning approach. ViTAcc \cite{li2022autovitacc} and \cite{dong2023heatvit} use int4 or int8 to represent the weights and activations. In contrast, our work is the first algorithm-hardware codesign to combine two pruning approaches. In terms of latency, our accelerator achieves $6.2-18.5\times$ latency reduction compared with the prior accelerator. As different accelerators use different numbers of computation units, which directly influences their peak performance (shown in Table \ref{tab:platform-specs}), we further normalize the latency by their respective peak performance ($\text{Normalized Latency}= \text{Latency}\times \text{Peak Performance}$) to obtain a fair comparison. Our accelerator achieves a normalized speedup of $1.5-4.5\times$ compared with SPViT \cite{kong2022spvit} and achieves a normalized speedup of $0.72-2.1\times$ compared with  HeatViT \cite{dong2023heatvit}. Our accelerator achieves higher speedup by executing the model with a higher pruning ratio. The achieved speedup is attributed to (1) in addition to token pruning, we further utilize the model pruning to reduce the computational complexity compared with \cite{kong2022spvit, dong2023heatvit}, (2) our architecture design using MPCA can efficiently utilize the block-wise data sparsity in the pruned model.

\begin{table}[]
\centering
\caption{Comparison with state-of-the-art ViT Accelerators}
\vspace{-0.3cm}
\label{tab:hardware-baselines}
\begin{adjustbox}{max width=0.48\textwidth}
\begin{tabular}{ccccc}
\toprule
& \textbf{ViTAcc} \cite{li2022autovitacc} & \textbf{HeatViT} \cite{dong2023heatvit} & \textbf{SPViT} \cite{kong2022spvit} & \textbf{Our Work} \\ 
\midrule
\midrule
\textbf{Platform} & \begin{tabular}[|c|]{@{}c@{}} Xilinx \\ ZCU102 \end{tabular}  & \begin{tabular}[|c|]{@{}c@{}}  Xilinx \\ ZCU102 \end{tabular}  & \begin{tabular}[|c|]{@{}c@{}}  Xilinx \\ ZCU102 \end{tabular}  & \begin{tabular}[|c|]{@{}c@{}}  Xilinx \\ Alveo U250 \end{tabular} \\ \midrule 
\textbf{Accuracy} & 77.94\% & 79.00\% & 79.34\% & 66.8\%-76.5\% \\  \midrule
\textbf{Quantization (bits)} & int4-8 & int8 & int16 & int16  \\ \midrule % I am using int16 for generating p_t p_h p_pe and estimate resource, usually, int 16 will not drop accuracy
\textbf{Model Pruning} & \xmark & \xmark & \xmark & \cmark \\ \midrule  
\textbf{Token Pruning} & \xmark & \cmark & \cmark & \cmark \\ \midrule  
\textbf{Latency(ms)} & 26 & 9.1-17.5 & 13.23 & 0.868-2.59\\ 
% \begin{tabular}[|c|]{@{}c@{}} \textbf{Computation} \\ \textbf{(GFLOPs)} \end{tabular} & N/A & 4.02 & 2.64 & \\  
\hline
\end{tabular}
\end{adjustbox}
\vspace{-0.27cm}
\end{table}

% \subsection{Inference Latency}

%\textbf{2-3 pages}

\section{Conclusion and Future Work}
\label{sec: disc}
In this paper, we proposed an algorithm-hardware codesign that simultaneously utilizes the \emph{static} weight pruning and \emph{dynamic} token pruning approaches. It bridges the gap of prior works that utilize only one pruning algorithm, further reducing the computation complexity of ViT. The proposed hardware accelerator can efficiently execute the pruned model through novel hardware architecture design. In the future, we plan to develop a design automation framework that automatically generates optimized implementation for the pruned ViT model given a target FPGA platform.


\section*{Acknowledgement}
This work is supported by the DEVCOM Army Research Lab (ARL) under grants W911NF2220159, and the National Science Foundation (NSF) under grants CCF-1919289 and SaTC-2104264. Equipment and support by AMD AECG are greatly appreciated.
\textbf{Distribution Statement A}: Approved for public release. Distribution is unlimited.

\bibliographystyle{IEEEtran}
\bibliography{IEEEabrv,main}

\end{document}